\documentclass[sigconf]{acmart}

\makeatletter                   
\def\mdseries@tt{m}             
\makeatother                    
\usepackage[draft=true]{minted} 
\usepackage{color}
\usepackage{hyperref}           
\hypersetup{
    colorlinks=false,
    linkcolor=blue,
    filecolor=red,      
    urlcolor=magenta,
    breaklinks=true,            
}
\usepackage{breakurl}           

\usepackage{tugraz_defaults}
\usepackage{enumitem}
\begin{document}
\sloppy                         

\title{NetSpectre: Read Arbitrary Memory over Network}
\renewcommand{\shorttitle}{NetSpectre}

\newcommand{\leakGadget}{\emph{leak gadget}\xspace}
\newcommand{\leakGadgets}{\emph{leak gadgets}\xspace} 
\newcommand{\transmitGadget}{\emph{transmit gadget}\xspace}
\newcommand{\transmitGadgets}{\emph{transmit gadgets}\xspace} 
\newcommand{\netGadget}{\emph{NetSpectre gadget}\xspace} 
\newcommand{\netGadgets}{\emph{NetSpectre gadgets}\xspace} 

\author{Michael Schwarz}
\affiliation{%
  \institution{Graz University of Technology}
}

\author{Martin Schwarzl}
\affiliation{%
  \institution{Graz University of Technology}
}

\author{Moritz Lipp}
\affiliation{%
  \institution{Graz University of Technology}
}

\author{Daniel Gruss}
\affiliation{%
  \institution{Graz University of Technology}
}

\renewcommand{\shortauthors}{M. Schwarz~\etal}

\acmDOI{XX.XXX/XXX_X}

\acmISBN{XXX-XXXX-XX-XXX/XX/XX}

\acmConference[Online]{arXiv}{July 2018}{}
\acmYear{2018}
\copyrightyear{2018}

\settopmatter{printacmref=false}

\begin{abstract}
Speculative execution is a crucial cornerstone to the performance of modern processors.
During speculative execution, the processor may perform operations the program usually would not perform.
While the architectural effects and results of such operations are discarded if the speculative execution is aborted, microarchitectural side effects may remain.
The recently published Spectre attacks exploit these side effects to read memory contents of other programs.
However, Spectre attacks require some form of local code execution on the target system.
Hence, systems where an attacker cannot run any code at all were, until now, thought to be safe.

In this paper, we present NetSpectre, a generic remote Spectre variant 1 attack. 
For this purpose, we demonstrate the first access-driven remote \EvictReload cache attack over network, leaking 15 bits per hour.
Beyond retrofitting existing attacks to a network scenario, we also demonstrate the first Spectre attack which does not use a cache covert channel.
Instead, we present a novel high-performance AVX-based covert channel that we use in our cache-free Spectre attack.
We show that in particular remote Spectre attacks perform significantly better with the AVX-based covert channel, leaking 60 bits per hour from the target system.
We verified that our NetSpectre attacks work in local-area networks as well as between virtual machines in the Google cloud.

NetSpectre marks a paradigm shift from local attacks, to remote attacks, exposing a much wider range and larger number of devices to Spectre attacks.
Spectre attacks now must also be considered on devices which do not run any potentially attacker-controlled code at all.
We show that especially in this remote scenario, attacks based on weaker gadgets which do not leak actual data, are still very powerful to break address-space layout randomization remotely.
Several of the Spectre gadgets we discuss are more versatile than anticipated.
In particular, value-thresholding is a technique we devise, which leaks a secret value without the typical bit selection mechanisms.
We outline challenges for future research on Spectre attacks and Spectre mitigations.
\end{abstract}

\maketitle

\vspace{1em}
\paragraph{\textbf{Responsible Disclosure}}
\textbf{
We disclosed our results to Intel on March 20th, 2018 and agreed on a disclosure date in late July 2018. 
}
\newpage
\section{Introduction}
Modern computers are highly optimized for performance.
However, these optimizations typically have side effects.
Side-channel attacks observe these side effects and consequently deduce information which would usually not be accessible to the attacker.
Software-based side-channel attacks are particularly unsettling since they do not require physical access to the device.
Many of these attacks fall into the category of microarchitectural attacks, which exploit differences in the timing or the behavior, which are caused by microarchitectural elements.

Over the past 20 years, software-based microarchitectural attacks have evolved from theoretical attacks~\cite{Kocher1996} on implementations of cryptographic algorithms~\cite{Osvik2006}, to more generic practical attacks~\cite{Yarom2014,Gruss2015Template}, and most recently to high potential threats~\cite{Lipp2018meltdown,Kocher2018} breaking the fundamental memory and process isolation.
Spectre~\cite{Kocher2018} is a microarchitectural attack, tricking another program into speculatively executing an instruction sequence which leaves microarchitectural side effects.
These side effects, in the case of all Spectre attacks demonstrated so far~\cite{Kocher2018,Trippel2018MeltdownPrime,Chen2018SGXpectre,Maisuradze2018speculose}, are timing differences caused by the pollution of data caches, \ie a traditional cache covert channel~\cite{Maurice2015C5,Liu2015}.

Speculative execution, which is used in Spectre attacks, is a crucial cornerstone to the performance of modern processors.
The branch prediction unit in modern processors makes an educated guess about which branch is taken and the processor then speculatively executes the expected instruction sequence following the predicted direction of the branch.
By manipulating the branch prediction, Spectre tricks a target process into performing a sequence of memory accesses which leak secrets from chosen virtual memory locations to the attacker.
This completely breaks confidentiality and renders virtually all security mechanisms on an affected system ineffective.
Spectre variant 1 is the Spectre variant which affects the largest number of devices, mostly associated with misspeculation following bound checks.
A code fragment performing first an operation such as a bound check and subsequently an operation with a microarchitectural side effect is called a ``Spectre gadget''.

Spectre attacks have so far been demonstrated in JavaScript~\cite{Kocher2018} and in native code~\cite{Kocher2018,Trippel2018MeltdownPrime,Chen2018SGXpectre,Maisuradze2018speculose}, but it is likely that any environment allowing sufficiently accurate timing measurements and some form of code execution enables these attacks.
Attacks on Intel SGX enclaves showed that enclaves are also vulnerable to Spectre attacks~\cite{Chen2018SGXpectre}.
However, there are billions of devices which never run any attacker-controlled code, \ie no JavaScript, no native code, and no other form of code execution on the target system.
Until now, these systems were believed to be safe against such attacks.
In fact, vendors are convinced that these systems are still safe and recommended to not take any action on these devices~\cite{Intel2018microrevguide}.

In this paper, we present NetSpectre, a new attack based on Spectre variant 1, requiring no attacker-controlled code on the target device, thus affecting billions of devices.
Similar to a local Spectre attack, our remote attack requires the presence of a Spectre gadget in the code of the target.
We show that systems containing the required Spectre gadgets in an exposed network interface or API can be attacked with our generic remote Spectre attack, allowing to read arbitrary memory over the network.
The attacker only sends a series of crafted requests to the victim and measures the response time to leak a secret value from the victim's memory.

We show that memory access latency, in general, can be reflected in the latency of network requests.
Hence, we demonstrate that it is possible for an attacker to distinguish cache hits and misses on specific cache lines remotely, by measuring and averaging over a larger number of measurements.
Based on this, we implemented the first access-driven remote cache attack, a remote variant of \EvictReload called \emph{Thrash+Reload}.
Our remote \emph{Thrash+Reload} attack is a significant leap forward from previous remote cache timing attacks on cryptographic algorithms~\cite{Aciicmez2007d,Jayasinghe2010remote,Aly2013attacking,Bernstein2005,Zhao2009cache,Cock2014timing}.
We facilitate this technique to retrofit existing Spectre attacks to our network-based scenario.
This NetSpectre variant is able to leak 15 bits per hour from a vulnerable target system.

By utilizing a previously unknown side channel based on the execution time of AVX2 instructions, we also demonstrate the first Spectre attack which does not rely on a cache covert channel at all.
Our AVX-based covert channel achieves a native code performance of 125 bytes per second at an error rate of \SI{0.58}{\percent}.
By using this covert channel in our NetSpectre attack instead of the cache covert channel, we achieve higher performance. 
Since cache eviction is not necessary anymore, we increase the speed of leaking to 60 bits per hour from the target system in a local-area network.
In the Google cloud, we can leak around 3 bits per hour from another independent virtual machine.

We demonstrate that using previously ignored gadgets allows breaking address-space layout randomization in a remote attack.
Address-space layout randomization (ASLR) is a defense mechanism deployed on most systems today, randomizing virtually all addresses.
An attacker with local code execution can easily bypass ASLR since ASLR mostly aims at defending against remote attacks but not local attacks.
Hence, many weaker gadgets for Spectre attacks were ignored so far, since they do not allow leaking actual data, but only address information.
However, moving to a remote attack scenario, these weaker gadgets become very powerful.

Spectre gadgets can be more versatile than anticipated in previous work.
This not only becomes apparent with the weaker gadgets we use in our remote ASLR break but even more so with the value-thresholding technique we propose.
Value-thresholding does not use the typical bit selection and memory reference mechanics as seen in previous Spectre attacks.
Instead, value-thresholding exploits information leakage in comparisons directly, by using a divide-and-conquer approach similar to a binary search.

NetSpectre marks a paradigm shift from local attacks to remote attacks.
This significantly broadens the range and increases the number of affected devices.
In particular, Spectre attacks must also be considered a threat to the security of devices which do not run any untrusted attacker-controlled code.
This shows that countermeasures must also be applied to these devices, which were previously thought to be safe.
We propose a new alternative to Retpolines~\cite{Turner2018retpoline} which has a clearer structure.
Future research on Spectre attacks and Spectre mitigations faces a series of challenges that we outline.
These challenges indicate that the current defenses can only be temporary solutions since they only fix symptoms without addressing the root cause of the problem.

\subheading{Contributions.} The contributions of this work are:
\begin{compactenum}
  \item We present NetSpectre, a generic remote Spectre variant 1 attack. For this purpose, we demonstrate the first access-driven remote cache attack (\EvictReload) over network, as a building block of NetSpectre.
  \item We demonstrate the first Spectre attack which does not utilize the cache.
  Instead, we present a new high-performance AVX-based covert channel which  significantly improves the performance of remote Spectre attacks.
  \item We show that even weaker forms of local Spectre attacks, which are incapable of leaking actual data, are still very powerful in remote Spectre attacks enabling remote ASLR breaks without any code execution on the device.
  \item We show that Spectre gadgets can be more versatile than anticipated. Our technique \emph{value-thresholding} allows obtaining a secret value without the typical bit selection and memory reference mechanics.
\end{compactenum}

\subheading{Outline.}
The remainder of the paper is organized as follows. 
In Section~\ref{sec:background}, we provide background on speculative execution and microarchitectural attacks.
In Section~\ref{sec:overview}, we provide an overview of the full NetSpectre attack.
In Section~\ref{sec:covert_channels}, we show how to build remote microarchitectural covert channels for use in NetSpectre attacks.
In Section~\ref{sec:attack}, we show how the building blocks are combined to extract memory contents over the network.
In Section~\ref{sec:evaluation}, we evaluate the performance of our attack.
In Section~\ref{sec:countermeasures}, we discuss countermeasures against local and network-based Spectre attacks and outline challenges for future research.
We conclude in Section~\ref{sec:conclusion}.

\section{Background}\label{sec:background}
In this section, we discuss out-of-order execution and a subset of out-of-order execution called speculative execution.
We detail branch prediction, a building block of most speculative execution implementations.
Finally, we will discuss known microarchitectural side-channel attacks as well as SIMD instructions as a better alternative for our use case.

\subsection{Out-of-order and Speculative Execution}\label{s:bg:specex}
Modern processors do not strictly execute one instruction after another.
Instead, modern processors have multiple execution units operating in parallel.
The serial instruction stream is distributed over these execution units, leaving fewer processor resources unused.
To retain the architecturally defined execution order, the processor has a so-called reorder buffer, which buffers operations until they are ready to be retired (made visible on the architectural level) in the order defined by the instruction stream.
Hence, out-of-order execution lets the processor precompute the results and effects of instructions.
Like simple pipeline processors, out-of-order processors suffer from interrupts, since any precomputed results and effects have to be discarded.
However, this is restricted to the architecturally visible state, and differences in the microarchitectural state may occur.
Instructions which get precomputed but not retired are called transient instructions~\cite{Kocher2018,Lipp2018meltdown}.

Out-of-order execution on modern processors can typically run several hundred simple instructions ahead of the architecturally visible state.
The actual number depends on the specific instructions and the size of the reorder buffer on the specific processor.

The instruction stream of almost every complex software is not purely linear but contains (conditional) branches.
Consequently, the processor often does not know which direction of the branch to follow ahead of time, \ie which subsequent instructions to run out of order.
In such cases, the processor uses prediction mechanisms to \emph{speculatively} execute instructions along one of the paths.
Hence, \emph{speculative execution} is a strict subset of out-of-order execution.
Correct predictions improve the performance and efficiency of the processor.
Incorrect predictions require discarding any precomputed results and effects after the misprediction.

\subsection{Branch Prediction}\label{s:branchprediction}\label{bg:bp}
Branch prediction is the most common prediction mechanism causing speculative execution.
Naturally, the performance and efficiency increase with the quality of the prediction.
Therefore, modern processors incorporate a number of branch prediction mechanisms. 

Intel~\cite{Intel_opt} processors have prediction mechanisms for ``Direct Calls and Jumps'', ``Indirect Calls and Jumps'', and ``Conditional Branches''.
These prediction mechanisms are implemented in different processor components, \eg the Branch Target Buffer (BTB)~\cite{Lee2017BranchShadowing,Evtyushkin2016ASLR}, the Branch History Buffer (BHB)~\cite{Bhattacharya2017perf}, and the Return Stack Buffer (RSB)~\cite{Fog2016}.
These buffers may be used in conjunction to obtain a good prediction~\cite{Fog2016}.
Since branch-prediction logic is typically not shared across physical cores~\cite{Ge2016}, the processor only learns from previous branches on the same core.

\subsection{Microarchitectural Attacks}
Most microarchitectural optimizations depend on the processed data or its location.
As a consequence, observing the effect of an optimization (\eg faster execution time) leaks information, \eg about the data or its location.

Traditionally, microarchitectural attacks were split into two categories:
side-channel attacks, which are non-destructive (passive), and fault attacks, which are destructive (active).
Side-channel attacks are often used to build covert channels, \ie a communication between two colluding parties through a side channel.

Microarchitectural side-channel attacks were first explored for attacks on cryptographic algorithms~\cite{Kocher1996,Tsunoo2003,Percival2005,Osvik2006,Gullasch2011}.
More recently, generic practical attack techniques were developed and used against a wide range of attack targets, \eg Flush+Reload~\cite{Gullasch2011,Yarom2014,Gruss2015Template}.

Microarchitectural attacks are usually considered software-based attacks, as opposed to traditional side-channel attacks and fault attacks requiring physical access to the device.
The most prominent example of a microarchitectural fault attack is Rowhammer, a hardware flaw in modern DRAM.
Rowhammer enables modification of privileged DRAM memory locations by an unprivileged attacker.

Meltdown~\cite{Lipp2018meltdown} and Spectre~\cite{Kocher2018} are two recent microarchitectural attacks.
They both use covert channels to transmit secrets, but the attacks themselves are no side-channel attacks.
Since they are non-destructive, they appear to fall in neither of the two categories.

Meltdown~\cite{Lipp2018meltdown} is a vulnerability present in many modern processors.
It is the basis of a series of attacks, which all bypass the isolation provided by the \texttt{user\_accessible} page-table bit (set to zero for kernel pages), \eg different attacks on KASLR~\cite{Jang2016,Gruss2016Prefetch,Hund2013} that were discovered independently before Meltdown~\cite{Lipp2018meltdown}.
The full Meltdown attack allows reading arbitrary kernel memory~\cite{Lipp2018meltdown}.

Spectre attacks~\cite{Kocher2018} exploit speculative execution, which is present in most modern processors.
Hence, they do not rely on any vulnerability, but solely on optimizations.
Through manipulation of the branch prediction mechanisms, an attacker lures a victim process into executing attacker-chosen code gadgets.
This enables the attacker to establish a covert channel from the speculative execution in the victim process to a receiver process under attacker control.

\subsection{Cache Attacks}
The largest group of microarchitectural attacks are cache attacks.
Cache attacks exploit timing differences introduced by small memory buffers, called caches. 
These CPU caches hide memory access latencies by buffering frequently used data in small but fast in-processor memories.
Modern CPUs have multiple cache levels that are either private per core or shared across cores.

Cache side-channel attacks were the first microarchitectural attacks.
Different cache attack techniques have been proposed in the past, including \EvictTime~\cite{Osvik2006}, \PrimeProbe~\cite{Osvik2006,Percival2005}, and \FlushReload~\cite{Yarom2014}.
Variations of these attacks are for instance \EvictReload~\cite{Gruss2015Template,Lipp2016}, and \FlushFlush~\cite{Gruss2016Flush}.
\FlushReload attacks and its variants work on a cache-line granularity, as they rely on shared memory.
Any cache line in shared memory will be a shared cache line in the inclusive last-level cache.
In a \FlushReload attack, the attacker constantly flushes a targeted memory location and measures the time it takes to reload the data.
If the reload time is low, the attacker learns that another process has loaded the cache line into the cache.
Various \FlushReload attacks have been demonstrated, \eg attacks on cryptographic algorithms~\cite{Yarom2014,Irazoqui2014,Benger2014}, web server function calls~\cite{Zhang2014}, specific system activity~\cite{Zhang2016ROPFR}, user input~\cite{Gruss2015Template,Lipp2016,Schwarz2018KeyDrown}, and kernel addressing information~\cite{Gruss2016Prefetch}.
\PrimeProbe follows a similar principle, but only has a cache-set granularity.
It works by occupying memory addresses and measuring when they are evicted from the cache.
Hence, \PrimeProbe attacks do not require any shared memory.
Various \PrimeProbe attacks have been demonstrated, \eg attacks on cryptographic algorithms~\cite{Osvik2006,Percival2005,Liu2015,Irazoqui2015SA}, user input~\cite{Lipp2016,Schwarz2018KeyDrown}, and kernel addressing information~\cite{Hund2013}.

Cache timing side channels have also been demonstrated in remote timing attacks.
\citeA{Bernstein2005} proposed a remote timing attack against a weak implementation of the AES algorithm.
The underlying timing difference was introduced from internal collisions in the algorithm and corresponding to that, a varying number of cache misses during AES computations.
Subsequently, several works were published improving and reproducing this attack~\cite{Aciicmez2007d,Jayasinghe2010remote,Aly2013attacking,Zhao2009cache}.

A special use case of a side-channel attack is a covert channel.
Here, the attacker controls both, the part that induces the side effect, and the part that measures the side effect.
This can be used to leak information from one security domain to another while bypassing any boundaries existing on the architectural level or above.
Both \PrimeProbe and \FlushReload have been used in high-performance covert channels~\cite{Liu2015,Maurice2017Hello,Gruss2016Flush}.
Meltdown~\cite{Lipp2018meltdown} and Spectre~\cite{Kocher2018} internally use a covert channel to transmit data from the transient execution to a persistent state.

\subsection{SIMD Instructions}
SIMD (single instruction multiple data) instructions allow performing an operation in parallel on multiple data values. 
SIMD instructions are available as instruction set extensions on a wide range of modern processors, \eg the Intel MMX extension~\cite{Intel_vol1,Intel_vol2,Intel_vol3,Peleg1996mmx}, the AMD 3DNow! extension~\cite{AMD64_manual,Oberman1999amd}, and the ARM VFP and NEON extensions~\cite{arm_arch_manualv8,arm_arch_manualv7,ARMGuide2002realview}.
On Intel, some of the SIMD instructions are processed by a dedicated SIMD unit within the processor core.
However, to avoid wasting energy, the SIMD unit is turned off when it is not used.
Consequently, to execute such SIMD instructions, the SIMD unit is first powered up, introducing a small latency on the first few SIMD instructions~\cite{Fog2016}.
Liu~\cite{Liu2017demand} mentioned that some SIMD instructions can be used to improve bus-contention covert channels since they enable a more direct access the memory bus.
However, so far, SIMD streaming instructions have not yet been used for pure SIMD covert channels or side-channel attacks.

\subsection{Advanced Persistent Threats}

The ever-increasing complexity in modern hardware and software is also reflected in modern malware.
Especially targeted malware like Stuxnet~\cite{Langner2011stuxnet}, Duqu~\cite{Bencsath2012duqu}, or Flame~\cite{Kushner2013real}, has proven to be extremely difficult to detect.
Consequently, it can persist on a target system or network over periods of weeks or months.
Hence, such malware is also known as ``advanced persistent threats'' (APTs)~\cite{Tankard2011advanced}.
In this scenario, also slow covert channels that transmit a few bits to bytes per day, \eg air-gap covert channels~\cite{Guri2014airhopper,Guri2015bitwhisper}, are highly practical, since they may run over a very long time.
APTs are usually a combination of a set of concrete exploits, bypassing different security mechanisms to achieve an overall goal.

\subsection{Address-Space Layout Randomization}
One security mechanism present in modern operating systems is address-space layout randomization (ASLR)~\cite{PaxASLR}.
It randomizes the locations of objects or regions in memory, \eg heap objects and stacks, so that an attacker cannot predict correct addresses. 
Naturally, this is a probabilistic approach, but it provides a significant gain in security in practice.
ASLR especially aims at mitigating control-flow-hijacking attacks, but it also makes other remote attacks difficult where the attacker has to provide a specific address.

\section{Attack Overview}\label{sec:overview}
In this section, we overview the NetSpectre attack based on a simple example. 
The building blocks of a NetSpectre attack are two \netGadgets: a \leakGadget, and a \transmitGadget.
We discuss the roles of these gadgets, which allow an attacker to perform a Spectre attack without any local code execution or access.
We discuss \netGadgets in detail, based on their type (leak or transmit) and the microarchitectural element they use (\eg cache).

Spectre attacks induce a victim to speculatively perform operations that would not occur during strictly serialized in-order processing of the program's instructions, and which leak a victim's confidential information via a covert channel to an attacker.
Spectre variant 1 induces speculative execution in the victim by mistraining a conditional branch, \eg a bounds check.
Spectre variant 2 induces speculative execution in the victim by maliciously injecting addresses into the branch-target buffer.
Although our attack can utilize any Spectre variant, we focus on Spectre variant 1 as it is the most widespread.
Moreover, according to Intel, in contrast to Meltdown and Spectre variant 2, variant 1 will not be fixed in hardware for the upcoming CPU generation~\cite{Intel2018spectrefix}. 

Before the actual condition is known, the CPU predicts the most likely outcome of the condition and then continues with the corresponding code path. 
There are several reasons why the result of the condition is not known at the time of evaluation, \eg a cache miss on parts of the condition, complex dependencies which are not yet satisfied, or a bottleneck in a required execution unit. 
By hiding these latencies, speculative execution leads to faster overall execution if the condition was predicted correctly. 
Intermediate results of a wrongly predicted condition are simply not committed to the architectural state, and the effective performance is similar as if the processor would never have performed any speculative execution.
However, any modifications of the microarchitectural state that occurred during speculative execution, such as the cache state, are not reverted. 

As our NetSpectre attack is mounted over the network, the victim device requires a network interface an attacker can reach.
The attacker must be able to send a large number of network packets to the victim.
However, these do not necessarily have to be within a short time frame.
Furthermore, the content of the packets in our attack is not required to be attacker-controlled.

In contrast to local Spectre attacks, our NetSpectre attack is not split into two phases.
Instead, the attacker constantly performs operations to mistrain the processor, which will make it constantly run into exploitably erroneous speculative execution.
NetSpectre does not mistrain across process boundaries, but instead trains in-place by passing valid and invalid values alternatingly to the exposed interface, \eg valid and invalid network packets.
For our NetSpectre attack, the attacker requires two Spectre gadgets, which are executed if a network packet is received:
a \leakGadget, and a \transmitGadget.
The \leakGadget accesses a bit stream at an attacker-controlled index, and changes some microarchitectural state depending on the state of the accessed bit. 
The \transmitGadget performs an arbitrary operation where the runtime depends on the microarchitectural state modified by the \leakGadget. 
Hidden in a significant amount of noise, the attacker can observe this timing difference in the network packet response time.
Spectre gadgets are commonly found in modern network drivers, network stacks, and network service implementation. 

To illustrate the working principle of our NetSpectre attack, we consider a basic example similar to the original Spectre variant 1 example~\cite{Kocher2018} in an adapted scenario: the code in \cref{lst:basic_example} is part of a function that is executed when a network packet is received.
We assume that \texttt{x} is attacker-controlled, \eg a field in a packet header or an index for some API.
This code forms our \leakGadget. 

\vspace{1em}
\begin{lstlisting}[caption={The conditional branch is part of a function executed when a network packet is processed.},label={lst:basic_example},language=C,style=customc,numbers=none]
if (x < bitstream_length)
   if(bitstream[x]) 
      flag = true
\end{lstlisting}

The code fragment begins with a bound check on \texttt{x}, a best practice when developing secure software. 
In particular, this check prevents the processor from reading sensitive memory outside of \texttt{bitstream}. 
Otherwise, an out-of-bounds input \texttt{x} could trigger an exception or could cause the processor to access sensitive memory by supplying $\texttt{x} = ($address of a secret bit to read$) - ($base address of \texttt{bitstream}$)$.  

To exploit the microarchitectural state change during speculative execution in a remote attack, the attacker has to adapt the original Spectre attack.
The attacker can remotely induce speculative execution as follows:
\begin{enumerate}[nolistsep, align=left, leftmargin=25pt, labelwidth=0pt, itemindent=-10pt]
 \item The attacker sends multiple network packets such that the attacker-chosen value of \texttt{x} is always in bounds.
 This trains the branch predictor, increasing the chance that the branch predictor predicts the outcome of the comparison as true.
 \item The attacker sends a packet where \texttt{x} is out of bounds, such that \texttt{bitstream[x]} is a secret bit in the target's memory. 
 \item Based on recent branch results of the condition, the branch predictor assumes the bounds check to be true, and the memory access is speculatively executed. 
\end{enumerate}

While changes in the architectural state are not committed after the correct result of the condition is resolved, changes in the microarchitectural state are not reverted. 
In the code in \cref{lst:basic_example} this means, that although the value of \texttt{flag} does not change, the cache state of \texttt{flag} does change. 
Only if the secret bit at \texttt{bitstream[x]} is set, \texttt{flag} is cached. 

The \transmitGadget is much simpler, as it only has to use \texttt{flag} in an arbitrary operation.
Consequently, the execution time of the gadget will depend on the cache state of \texttt{flag}. 
In the most simple case, the \transmitGadget simply returns the value of \texttt{flag}, which is set by the \leakGadget. 
As the architectural state of \texttt{flag} (\ie its value) does not change for out-of-bounds \texttt{x}, it does not leak secret information. 
However, the response time of the \transmitGadget depends on the microarchitectural state of \texttt{flag} (\ie whether it is cached), which does leak a secret bit. 

To complete the attack, the attacker measures the response time for every secret bit to leak.
As the difference in the response time is in the range of nanoseconds, the attacker needs to average over a large number of measurements to obtain the secret value with acceptable confidence.
Indeed, our experiments show that the difference in the microarchitectural state becomes visible when performing a large number of measurements.
Hence, an attacker can first measure the two corner cases (\ie cached and uncached) and afterward, to extract a real secret bit, perform as many measurements as necessary to distinguish which case it is with sufficient confidence, \eg using a threshold or a Bayes classifier.

\begin{figureA}[t]{gadget_types}[The interaction of the \netGadget types.]
\resizebox{!}{3.5cm}{
\begin{tikzpicture}[yscale=0.8]

\draw (0.5,0) rectangle +(0.5,0.5) node[pos=.5] {0};
\draw (1,0) rectangle +(0.5,0.5) node[pos=.5] {1};
\draw (1.5,0) rectangle +(0.5,0.5) node[pos=.5] {0};
\draw (2,0) rectangle +(0.5,0.5) node[pos=.5] {1};

\draw[fill=red] (2.5,0) rectangle +(0.5,0.5) node[pos=.5,white] {0};
\draw[fill=red] (3,0) rectangle +(0.5,0.5) node[pos=.5,white] {0};
\draw[fill=red] (3.5,0) rectangle +(0.5,0.5) node[pos=.5,white] {0};

\node at (1.5,-0.2) {\scriptsize bitstream};
\node at (3.25,-0.2) {\scriptsize (out of bounds)};

\draw[densely dotted,thick] (2,1.25) rectangle +(3,0.75) node[pos=.5] {Leak Gadget};
\draw[fill=gray!20] (3.5,-1.5) rectangle +(1.75,1) node[pos=.5,text width=3.5em] {$\mu$-arch. Element};
\draw[densely dotted,thick] (0.5,-1.38) rectangle +(3,0.75) node[pos=.5] {Transmit Gadget};

\draw[>=stealth,->,in=270,out=90,thick] (2.75,0.5) to node[above,near start,sloped] {leak} (3.5,1.25);
\draw[>=stealth,->,in=90,out=270,thick] (4,1.25) to node[above,midway,sloped] {encode} (4.75,-0.5);

\draw[>=stealth,->,in=180,out=90,thick] (0,-2.5) to node[midway,sloped,above] {index} (2,1.65);
\draw[>=stealth,->,thick] (1,-2.5) to (1,-1.38);
\draw[>=stealth,->,thick] (2.75,-1.38) to (2.75,-2.5);

\draw (-0.5,2.5) rectangle (6,-2);
\draw[densely dashed] (-0.5,-2.12) to (6,-2.12);
\node at (0.25,2.25) {\small Victim};
\node at (4.75,-2.35) {\small Network interface};

\node at (2,-2.75) {$\Delta$ = leaked bit};
\node at (0,-2.75) {bit index};
\end{tikzpicture}
}
\end{figureA}

We refer to the two gadgets, the \leakGadget and the \transmitGadget, as \netGadgets. 
Running a \netGadget may require sending more than one packet.
Furthermore, the \leakGadget and \transmitGadget may be reachable via different independent interfaces, \ie both interfaces must be accessible for the attacker. 
\Cref{fig:gadget_types} illustrates the two types of gadgets, which will be described in more detail later in this section. 

\subsection{Gadget location}
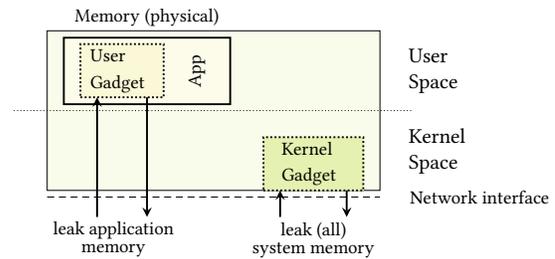
\begin{figureA}[t]{gadget_location}[Depending on the gadget location, the attacker has access to either the memory of the entire corresponding application or the entire kernel memory, typically including the entire system memory.]
\resizebox{!}{3.5cm}{
\begin{tikzpicture}[yscale=0.8]

\draw[fill=green!10!white] (0,0) rectangle +(5,3);
\draw[densely dotted] (-0.5,1.5) to (5.5,1.5);
\node[text width=2em] at (5.75, 0.75) {Kernel Space};
\node[text width=2em] at (5.75, 2.25) {User Space};

\node at (1.5, 3.25) {\small Memory (physical)};

\draw[fill=green!40!white,thick,densely dotted] (3.25,0) rectangle +(1.5,1) node[pos=.5,text width=3em] {\small Kernel Gadget};
\draw[densely dashed] (0,-0.125) to (5,-0.125);

\node at (6.5,-0.125) {\small Network interface};

\draw[->,>=stealth,thick] (3.5,-0.5) to (3.5,0);
\draw[->,>=stealth,thick] (4.5,0) to (4.5,-0.5);

\draw[fill=yellow!20!white,thick] (0.25,1.62) rectangle +(2.5,1.25);
\draw[fill=yellow!40!white,thick,densely dotted] (0.5,1.75) rectangle +(1.25,1) node[pos=.5,text width=3em] {\small User Gadget};
\node[rotate=90] at (2.25,2.25) {\small App};

\draw[->,>=stealth,thick] (0.75,-0.5) to (0.75,1.75);
\draw[->,>=stealth,thick] (1.5,1.75) to (1.5,-0.5);

\node at (1,-0.75) {\small leak application};
\node at (1,-1.12) {\small memory};

\node at (4,-0.75) {\small leak (all)};
\node at (4,-1.12) {\small system memory};

\end{tikzpicture}
}
\end{figureA}

The set of attack targets depends on the location of the \netGadgets.
As illustrated in \Cref{fig:gadget_location}, on a high level, there are two different gadget locations: gadgets are either located in the user space or in the kernel space.

\subsubsection{Attacks on the Kernel}
The network driver is usually implemented in the kernel of the operating system, either as a fixed component or as a kernel module.
In either case, kernel code is executed when a network packet is received. 
If any kernel code processed during the handling of the network packet contains a \netGadget, \ie an attacker-controlled part of the packet is used as an index to reference a bit, a NetSpectre attack is possible. 

An attack on the kernel code is particularly powerful, as the kernel does not only have the kernel memory mapped but typically also the entire physical memory. 
On Linux and macOS, the physical memory can be accessed via the direct-physical map, \ie every physical memory location is accessible via a predefined virtual address in the kernel address space. 
Windows does not use a direct-physical map but maintains memory pools, which typically also map a large fraction of the physical memory. 
Thus, a NetSpectre attack leveraging a \netGadget in the kernel can in general leak an arbitrary bit stored in memory. 

\subsubsection{Attacks on the User Space}
Usually, network packets are not only handled by the kernel but are also passed on to a user-space application which processes the content of the packet. 
Hence, not only the kernel but also user-space applications can contain \netGadgets. 
In fact, all code paths that are executed when a network packet arrives are candidates to look for \netGadgets. 
This does includes code both on the server side and the client side.

An advantage in attacking user-space applications is the significantly larger attack surface, as many applications process network packets. 
Especially on servers, there are an abundance of services processing user-controlled network packets, \eg web servers, FTP servers, or SSH daemons. 
But also, a remote server can attack a client machine, \eg via web sockets, or SSH connections.
In contrast to attacks on the kernel space, which in general can leak any data stored in the system memory, attacks on a user-space application can only leak secrets of the attacked application. 

Such application-specific secrets include secrets of the application itself, \eg credentials and keys.
Thus, a NetSpectre attack leveraging a \netGadget in an application can access arbitrary data processed by the application. 
Furthermore, if the victim is a multi-user application, \eg a web server, it also contains secrets of multiple users. 
Especially for popular websites, this can easily affect thousands or millions of users.

\subsection{Gadget type}
We now discuss the different \netGadgets, namely the \leakGadget to encode a secret bit into a microarchitectural state, and the \transmitGadget to transfer the microarchitectural state to a remote attacker.

\subsubsection{Leak Gadget} 
The first type of gadget, the \leakGadget, leaks secret data by changing a microarchitectural state depending on the value of a memory location that would not be accessible directly through any interface accessible to the attacker. 
Note that this state change happens on the victim device, and is not directly observable over the network. 

A \leakGadget gadget can leak a single bit, or even one or multiple bytes. 
Single-bit gadgets are the most versatile, as storing a one-bit (binary) state can be accomplished with many microarchitectural states, as only two cases have to be distinguished (\cf \cref{sec:covert_channels}). 
\citeA{Kocher2018} leaked the secret data with a byte-wise gadget. 
This simplifies the access to the secret data, as only byte indices have to be used, but complicates the recovery process, as 256 states have to be distinguished. 
With local Spectre attacks, the recovery process is implemented by the attacker, and thus a complex recovery process does not have any drawbacks but a slightly lower performance. 
The reason is that a larger number of side-channel tests (\eg more \FlushReload tests) have to be performed on the receiving side of the covert channel. 
\citeA{Lipp2018meltdown} showed that a transmission from out-of-order execution with single-bit covert channel can be significantly faster than a byte-wise or multi-byte covert channel in a similar attack.
NetSpectre attacks have to rely on gadgets for the recovery process, slowing down the transmission significantly.
A single-bit gadget does not only have several microarchitectural elements to choose from, but the data is also comparably easy to recover, and the data transmission is faster since fewer remote side-channel tests have to be performed for the covert channel transmission. 
Thus, we focus on single-bit \leakGadgets in this paper. 
Single-bit \leakGadgets can be as simple as shown in \cref{lst:basic_example}. 
In this example, a value (\texttt{flag}) is cached if the bit at the attacker-chosen location is set. 
If the bit is not set, the cache state of the variable remains unchanged, \ie if it was previously uncached, it will not be cached.
Hence, the attacker can use this gadget to leak secret bits into the microarchitectural state.

\subsubsection{Transmit Gadget}
In contrast to Spectre, NetSpectre requires an additional gadget to transmit the leaked information to the attacker. 
As the attacker does not control any code on the victim device, the recovery process, \ie converting the microarchitectural state back into an architectural state, cannot be implemented by the attacker.
Furthermore, the architectural state can usually not be accessed via the network and thus, it would not even help if the gadget converts a microarchitectural into an architectural state. 

From the attacker's perspective, the microarchitectural state must become visible over the network. 
This may not only happen directly via the content of a network packet but also via side effects. 
And indeed, the microarchitectural state will in some cases become visible to the attacker, \eg in the form of the response time.
We refer to a code fragment which exposes the microarchitectural state to a network-based attacker and which can be triggered by an attacker, as a \transmitGadget.
Naturally, the \transmitGadget has to be located on the victim device. 
With a \transmitGadget, the measurement of the microarchitectural state happens on a remote machine but exposes the microarchitectural over a network-reachable interface.

In the original Spectre attack, \FlushReload is used to transfer the microarchitectural state to an architectural state, which is then read by the attacker to leak the secret. 
The ideal case would be if such a \FlushReload gadget is available on the victim, and the architectural state can be observed over the network. 
However, as it is unlikely to locate an exploitable \FlushReload gadget on the victim and access the architectural state, regular Spectre gadgets cannot simply be retrofitted to mount a NetSpectre attack. 

In the most direct case, the microarchitectural state becomes visible for a remote attacker, through the latency of a network packet. 
A simple \transmitGadget for the \leakGadget shown in \cref{lst:basic_example} just accesses the variable \texttt{flag}. 
The response time of the network packet depends on the cache state of the variable, \ie if the variable was accessed, the response takes less time. 
Generally, an attacker can observe changes in the microarchitectural state if such differences are measurable via the network. 

\section{Remote Microarchitectural Covert Channels}\label{sec:covert_channels}
As described in the last section, a cornerstone of our NetSpectre attack is building a microarchitectural covert channel that exposes information to a remote attacker.
Since in our scenario, the attacker cannot run any code on the target system, we assume the transmission happens through a \transmitGadget whose execution can be triggered by the attacker.
In this section, we demonstrate the first remote access-driven cache attack, \emph{Thrash+Reload}, a variant of \EvictReload.
We show that with this remote cache attack, an attacker can build a covert channel from the speculative execution on the target device to a remote receiving end on the attacker's machine.
Furthermore, we also present a previously unknown microarchitectural covert channel based on AVX2 instructions.
We show that this covert channel can be used in NetSpectre attacks, yielding even higher transmission rates than the remote cache covert channel.

\subsection{Remote Cache Covert Channel}
\citeA{Kocher2018} leverage the cache as the microarchitectural element to encode the leaked data. 
This allows using well-known cache side-channel attacks, such as \FlushReload~\cite{Yarom2014} or \PrimeProbe~\cite{Osvik2006,Percival2005} to deduce the microarchitectural state and thus the encoded data. 

However, not only caches keep microarchitectural states which can be made visible on the architectural level.
Methods to extract the microarchitectural state from elements such as DRAM~\cite{Pessl2016}, BTB~\cite{Evtyushkin2016ASLR}, or RSB~\cite{Bulygin2008cpu} are known. 
Generally, the receiver of every microarchitectural covert channel~\cite{Ge2016} can be used to transfer a microarchitectural state to an architectural state.

Mounting a Spectre attack by leveraging the cache has three main advantages: there are powerful methods to make the cache state visible, many operations modify the cache state and are thus visible in the cache, and the timing difference between a cache hit and cache miss is comparably large.
\FlushReload is usually considered the most fine-grained and accurate cache attack, with almost zero noise~\cite{Yarom2014,Gruss2016Flush,Ge2016}.
If \FlushReload is not applicable in a certain scenario, \PrimeProbe is considered the next best choice~\cite{Maurice2017Hello,Schwarz2017MGX}.
Consequently, all Spectre attacks published so far use either \FlushReload~\cite{Kocher2018,Chen2018SGXpectre} or \PrimeProbe~\cite{Trippel2018MeltdownPrime}.

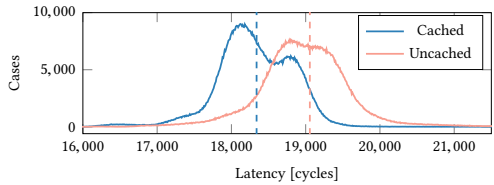
\begin{figureA}[t]{network_hist}[Measuring the response time of a simple \transmitGadget, that accesses a certain variable. Only by performing a large number of measurements, the difference in the response timings depending on the cache state of the variable becomes visible. The average values of the two distributions are shown as dashed vertical lines.]
\resizebox{!}{2.5cm}{
\begin{tikzpicture}
\begin{axis}[
style={font=\footnotesize},
xlabel={Latency [cycles]},
ylabel={Cases},
width=0.95\hsize,
scaled y ticks=false,
scaled x ticks=false,
height=3.5cm,
xmin=16000,
xmax=21500,
ymin=-600,
ymax=10000
]
\addplot+[blue,thick,mark=none] table[x=cycle,y=hit,col sep=comma] {data/network_6200u.csv};
\addplot+[red!50,thick,mark=none] table[x=cycle,y=miss,col sep=comma] {data/network_6200u.csv};
\legend{Cached,Uncached}
\addplot[thick, dashed, blue] coordinates {(18339,-700)(18339,25000)};
\addplot[thick, dashed, red!50] coordinates {(19056,-700)(19056,25000)};

\end{axis}
\end{tikzpicture}
}
\end{figureA}

To build our first NetSpectre attack, we need to adapt local cache covert channel techniques.
Instead of measuring the memory access time directly, we measure the response time of a network request which uses the corresponding memory location.
Consequently, the response time will be influenced by the cache state of the variable used for the attack.
The difference in the response time due to the cache state will be in the range of nanoseconds since memory accesses are comparably fast.

The network latency is subject to many factors, leading to noisy results.
However, the influence of noise can be decreased by averaging over a large amount of network packets~\cite{Aciicmez2007d,Jayasinghe2010remote,Aly2013attacking,Bernstein2005,Zhao2009cache}. 
Hence, an attacker needs to average over a large number of measurements to obtain the secret value with acceptable confidence.

\Cref{fig:network_hist} shows that the difference in the microarchitectural state is indeed visible when performing a large number of measurements.
The average values of the two distributions are illustrated as dashed vertical lines.
An attacker can either use a classifier on the measured values, or first measure the two corner cases (cached and uncached) to get a threshold for the real measurements. 

Still, as the measurement destroy the cache state, \ie the variable is always cached after the first measurement, the attacker requires a method to evict (or flush) the variable from the cache. 
As it is unlikely that the victim provides an interface to flush or evict a variable directly, the attacker cannot use well-known cache attacks but has to resort to more crude methods. 
Instead of the targeted eviction in \EvictReload, we simply evict the entire last-level cache by thrashing the cache, similar as \citeA{Maurice2015C5}. 
Hence, we call this technique \emph{Thrash+Reload}.
To thrash the entire cache without code execution, we again have to use an interface accessible via the network. 
In the simplest form, any packet sent from the victim to the attacker, \eg a file download, has the chance to evict the variable from the cache. 

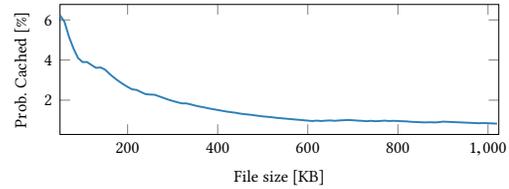
\begin{figureA}[t]{thrashreload}[The probability that a specific variable is evicted from the victim's last-level cache by downloading a file from the victim (Intel i5-6200U). The larger the downloaded file, the higher the probability that the variable is evicted.]
\resizebox{!}{2.5cm}{
\begin{tikzpicture}
\begin{axis}[
style={font=\footnotesize},
xlabel={File size [KB]},
ylabel={Prob. Cached [\%]},
width=0.95\hsize,
scaled y ticks=false,
height=3.5cm,
xmin=50,
xmax=1024
]
\addplot+[blue,thick,mark=none] table[x=size,y=cached,col sep=comma] {data/thrash1.csv};


\end{axis}
\end{tikzpicture}

}
\end{figureA}

\Cref{fig:thrashreload} shows the probability of evicting a specific variable (\ie the \texttt{flag} variable) from the last-level cache by requesting a file from the victim. 
The victim is running on an Intel i5-6200U with \SI{3}{\mega B} last-level cache. 
Downloading a file with 590 kilobytes is already sufficient to evict the variable with a probability of $\geq$ \SI{99}{\percent}. 

With a mechanism to distinguish cache hits and misses, as well as a mechanism to throw things out of the cache, we have all building blocks required for a cache side-channel attack or a cache covert channel.
\emph{Thrash+Reload} combines both mechanisms over a network interface, forming the first remote cache covert channel.
In our experiments on a local-area network, we achieve a transmission rate of up to 4 bit per minute, with an error rate of $<0.1\,\%$. 
This is significantly slower than cache covert channels in a local native environment, \eg the most similar attack (\EvictReload) achieves a performance of \SI{13.6}{\kilo b/s} with an error rate of \SI{3.79}{\percent}.

In this paper, we use our remote cache covert channel for remote Spectre attacks.
However, remote cache covert channels and especially remote cache side-channel attacks are an interesting object of study.
Many attacks that were presented previously would be devastating if mounted over a network interface~\cite{Yarom2014,Gruss2015Template,Gras2017AnC}.

\subsection{Remote AVX-based Covert Channel}\label{sec:avx-channel}
To demonstrate the first Spectre variant which does not rely on the cache as the microarchitectural element, we require a covert channel which allows transmitting information from speculative execution to an architectural state.
Thus, we build a novel covert channel based on timing differences in AVX2 instructions.
This covert channel has a low error rate and high performance, and it allows for a significant performance improvement in our NetSpectre attack as compared to the remote cache covert channel.

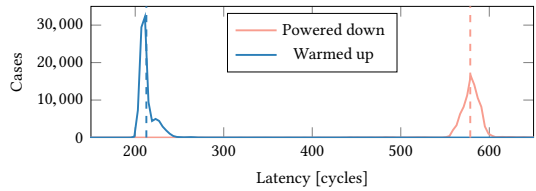
\begin{figureA}[t]{avx_hist}[Differences in the execution time for AVX2 instructions (Intel i5-6200U). If the AVX2 unit is inactive (powered down), executing a \SIx{256}-bit instruction takes on average \SIx{366} cycles longer than on an active AVX2 unit. The average values are shown as dashed vertical lines.]
\resizebox{!}{2.5cm}{
\begin{tikzpicture}
\begin{axis}[
style={font=\footnotesize},
xlabel={Latency [cycles]},
ylabel={Cases},
width=0.95\hsize,
scaled y ticks=false,
height=3.5cm,
ymin=0,
xmin=150,
xmax=650,
ymax=35000,
legend style={at={(0.5,0.95)},anchor=north}
]
\addplot+[red!50,thick,mark=none] table[x=cycle,y=miss,col sep=comma] {data/avx_hist.csv};
\addplot+[blue,thick,mark=none] table[x=cycle,y=hit,col sep=comma] {data/avx_hist.csv};

\legend{Powered down,Warmed up}
\addplot[thick, dashed, red!50] coordinates {(578.5,0)(578.5,35000)};
\addplot[thick, dashed, blue] coordinates {(212.6,0)(212.6,35000)};

\end{axis}
\end{tikzpicture}
}
\end{figureA}

To save power, the CPU can power down the upper half of the AVX2 unit which is used to perform operations on \SIx{256}-bit registers. 
The upper half of the unit is powered up as soon as an instruction is executed which uses \SIx{256}-bit values~\cite{Fog2015AVXa}. 
If the unit is not used for more than \SI{1}{\milli\second}, it is powered down again~\cite{Fog2015AVXb}. 

Performing a \SIx{256}-bit operation when the upper half is powered down incurs a significant performance penalty. 
For example, we measured the execution (including measurement overhead) of a simple bit-wise AND of two \SIx{256}-bit registers (\texttt{VPAND}) on an Intel i5-6200U (\cf \Cref{fig:avx_hist}).
If the upper half is active, the operation takes on average \SIx{210} cycles, whereas if the upper half is powered down (\ie it is inactive), the operation takes on average \SIx{576} cycles, resulting in a difference of \SIx{366} cycles. 
The difference is even larger than the difference between cache hits and misses, which is only \SIx{160} cycles on the same system. 
Hence, the timing difference in AVX2 instructions is better for remote microarchitectural attacks than the timing difference between cache hits and misses.

Similarly to the cache, reading the latency of an AVX2 instruction also destroys the encoded information. 
An attacker therefore requires a method to reset the AVX2 unit, \ie power down the upper half. 
In contrast to the cache, this is significantly easier, as the upper half of the AVX2 unit is automatically powered down after \SI{1}{\milli\second} of inactivity. 
Thus, an attacker only has to wait at least \SI{1}{\milli\second} before the next measurement. 

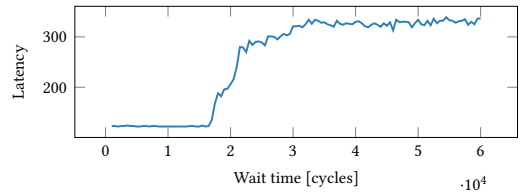
\begin{figureA}[t]{avx_powerdown}[The number of cycles it takes to execute the \texttt{VPAND} instruction (including measurement overhead) after not using the AVX2 unit. After approximately 0.5\,ms, the upper half of the AVX2 unit starts to power down, which increases the latency for subsequent AVX2 instructions. After approximately 1\,ms, it is fully powered down, and we see the maximum latency for subsequent AVX2 instructions.]
\resizebox{!}{2.5cm}{
\begin{tikzpicture}
\begin{axis}[
style={font=\footnotesize},
xlabel={Wait time [cycles]},
ylabel={Latency},
width=0.95\hsize,
scaled y ticks=false,
height=3.5cm
]
\addplot+[blue,thick,mark=none] table[x=cycle,y=latency,col sep=comma] {data/avx_powerdown.csv};


\end{axis}
\end{tikzpicture}

}
\end{figureA}

\Cref{fig:avx_powerdown} shows the execution time of a \SIx{256}-bit AVX2 instruction (specifically \texttt{VPAND}) after inactivity of the AVX2 unit. 
If the inactivity is shorter than \SI{0.5}{\milli\second}, \ie the last AVX2 instruction was executed not more than \SI{0.5}{\milli\second} ago, there is no performance penalty when executing an AVX2 instruction which uses the upper half of the AVX2 unit. 
After that, the AVX2 unit begins powering down, increasing the execution time for any subsequent AVX2 instruction, since the unit has to be powered up again and only emulates AVX2 in the meantime~\cite{Fog2015AVXb}. 
The AVX2 unit is fully powered down after approximately \SI{1}{\milli\second}, leading to the highest performance penalty if any AVX2 instruction is executed in this state. 

\begin{lstlisting}[caption={An AVX2 NetSpectre gadget which encodes a bit using a 256-bit instruction.},label={lst:avx2_leak},language=C,style=customc,numbers=none,float]
if (x < bitstream_length)
   if(bitstream[x]) 
      _mm256_instruction();
\end{lstlisting}

A \leakGadget leveraging AVX2 is similar to a \leakGadget leveraging the cache. 
\Cref{lst:avx2_leak} shows an example (pseudo-)code of an AVX2 \leakGadget. 
The \texttt{\_mm256\_instruction} represents an arbitrary \SIx{256}-bit AVX2 instruction, \eg \texttt{\_mm256\_and\_si256}. 
If the referenced bit \texttt{x} in the bit stream \texttt{bitstream} is set, the instruction is executed, and as a consequence, the upper half of the AVX2 unit is powered on.
This is also true if the branch-prediction outcome was not correct and the AVX2 instruction is accessed during speculative execution.
Note that there is no data dependency between the AVX2 instruction and either the bit stream or the index.
Only the information whether an AVX2 instruction was executed is used to transmit the secret bit through the covert channel.

The \transmitGadget is again similar to the \transmitGadget for the cache. 
Any function that uses an AVX2 instruction, and has thus a measurable runtime difference observable over the network, can be used as a \transmitGadget. 
Even the \leakGadget shown in \Cref{lst:avx2_leak} can act as a \transmitGadget. 
By providing an in-bounds value for \texttt{x}, the runtime of the function depends on the state of the upper half of the AVX2 unit. 
If the upper half of the unit was used before, \ie a `1'-bit was leaked, the function executes faster than if the upper half was not used before, \ie a `0'-bit was leaked. 

With these building blocks, we can build an AVX-based covert channel.
Our covert channel is the first pure-AVX covert channel and the first AVX-based remote covert channel.
In our experiments in a native local environment, we achieve a transmission rate of \SI{125}{B/s} with an error rate of \SI{0.58}{\percent}.
In a local-area network, we achieve a transmission rate of \SI{8}{B/\minute}, with an error rate of $<$\SI{0.1}{\percent}. 
Since the true capacity of this remote covert channel is higher than the true capacity of our remote cache covert channel, we can already see that it yields higher performance in our NetSpectre attack.

\section{Attack Variants}\label{sec:attack}

In this section, we describe two NetSpectre attack variants.
The first attack allows extracting secret data bit-by-bit from the memory of the target system.
The second attacks allows defeating ASLR on the remote machine, paving the way for remote exploitation of bugs that ASLR would usually mitigate.
We use gadgets based on Spectre variant 1 for illustrative purposes but this can naturally be done with any Spectre gadget that lies in a code path reached from handling of a remote packet.

\subsection{Extracting Data from the Target System}
With typical \netGadgets (\cf \cref{sec:overview}), the extraction process consists of 4 steps.
Note that depending on the gadgets, the \leakGadget and \transmitGadget might be the same. 
\begin{enumerate}[nolistsep, align=left, leftmargin=25pt, labelwidth=0pt, itemindent=-10pt]
 \item Mistrain the branch predictor.
 \item Reset the state of the microarchitectural element.
 \item Leak a bit to the microarchitectural element.
 \item Expose state of the microarchitectural element to the network.
\end{enumerate}

In step 1, the attacker mistrains the branch predictor of the victim to run a Spectre attack. 
To mistrain the branch predictor, the attacker leverages the \leakGadget with valid indices. 
The valid indices ensure that the branch predictor learns to always take the branch, \ie the branch predictor speculates that the condition is true. 
Note that this step only relies on the \leakGadget. 
There is no feedback to the attacker, and thus the microarchitectural state does not have to be reset or transmitted. 

In step 2, the attacker has to reset the microarchitectural state to enable the encoding of leaked bits using a microarchitectural element. 
This step highly depends on the used microarchitectural element, \eg when leveraging the cache, the attacker downloads a large file from the victim (\cf \Cref{fig:thrashreload}), if AVX2 is used, the attacker simply waits for more than \SI{1}{\milli\second}. 
After this step, all requirements are satisfied to leak a bit from the victim. 

In step 3, the attacker exploits the Spectre vulnerability to leak a single bit from the victim. 
As the branch predictor is mistrained in step 1, providing an out-of-bounds index to the \leakGadget will run the in-bounds path and modify the microarchitectural element, \ie the bit is encoded in the microarchitectural element.

In step 4, the attacker has to transmit the encoded information via the network. 
This step corresponds to the second phase of the original Spectre attack. 
In contrast to the original Spectre attack, which leverages a cache attack, the attacker uses the \transmitGadget for this step as described in \cref{sec:covert_channels}. 
The attacker sends a network packet which is handled by the \transmitGadget and measures the time from sending the packet until the response arrives. 
As described in \cref{sec:covert_channels}, this round-trip time depends on the state of the microarchitectural element, and thus on the leaked bit. 

As the network latency varies, the four steps have to be repeated multiple times to eliminate the noise caused by these fluctuations. 
Typically, the variance in latency follows a certain distribution depending on multiple factors, such as distance, number of hops, network congestion~\cite{Hopper2010Tor,Goonatilake2012Latency,Charneski2015}. 
The number of repetitions depends mainly on the variance in latency of the network connection. 
Thus, depending on the latency distribution, the number of repetitions can be deduced using statistical methods. 
In \Cref{sec:eval:results}, we evaluate this attack variant and provide empirically determined numbers for our attack setup.

\subsection{Remotely Breaking ASLR on the Target System}
If the attacker has no access to a bit-leaking \netGadgets, it is possible to use a weaker \netGadget which does not leak the actual data but only information about the corresponding address.
Such gadgets were not considered harmful for Spectre attacks, which already have local code execution, as ASLR does not protect against local attacks.
However, in a remote scenario, it is very valuable to break ASLR.
If such a \netGadget is found in a user-space program, it breaks ASLR for this process. 

\begin{lstlisting}[caption={A NetSpectre gadget which can be leveraged to break ASLR.},label={lst:aslr_gadget},language=C,style=customc,numbers=none,float]
if (x < array_length)
   access(array[x])
\end{lstlisting}

\Cref{lst:aslr_gadget} shows a simple \leakGadget which is already sufficient to break ASLR. 
With the help of this gadget, breaking ASLR consists of 3 steps.
\begin{enumerate}[nolistsep, align=left, leftmargin=25pt, labelwidth=0pt, itemindent=-10pt]
 \item Mistrain the branch predictor.
 \item Access an out-of-bounds index to cache a (known) memory location.
 \item Measure the execution time of a function via the network to deduce whether the out-of-bounds access cached a part of it. 
\end{enumerate}

The mistraining step is the same as for any Spectre attack, leading to speculative out-of-bounds accesses relative to the array. 
If the attacker provides an out-of-bounds value for \texttt{x} after mistraining, the array element at this index is speculatively accessed. 
Assuming a byte array and an (unsigned) 64-bit index, an attacker can (speculatively) access any memory location, as the index wraps around if the base address plus the index is larger than the virtual memory. 
If the byte at this memory location is valid and cacheable, it is cached after executing this gadget, as the speculative execution will fetch the corresponding memory location into the cache. 
Thus, this gadget allows caching arbitrary memory locations which are valid in the current virtual memory, \ie every mapped memory location of the current application. 

The attacker uses this gadget to cache a memory location at a known location, \eg the vsyscall page which is mapped into every application at the same virtual address~\cite{LWN2011vsyscall}.
The attacker then measures the execution time of a function accessing the now cached memory location, \eg older versions of \texttt{time} or \texttt{gettimeofday}. 
If the function executes faster, the out-of-bounds array index actually cached a memory location used by this function. 
Thus, from the known address and the value of the array index, \ie the relative offset to the known address, the attacker can calculate the actual address of the \leakGadget. 

With an ASLR entropy of \SI{30}{b} on Linux~\cite{Marco2016exploiting}, there are $2^{30}$ possible offsets the attacker has to check.
Due to the KPTI (formerly KAISER~\cite{Gruss2017KASLR}) patches, no other page close to the vsyscall page is mapped in the user space. 
Consequently, in the $2^{30}$ possible offsets, there is only a single valid, and thus cacheable, offset.
Hence, we can perform a binary search to find the correct offset, \ie speculatively try to load half of the possible offsets into the cache and check a single time.
If the single valid, and thus cacheable, offset was cached, the attacker chose the correct half, otherwise, the attacker continues with the other half.
This reduces the number of checks to defeat ASLR to only $30$.

Although vsyscall is a legacy feature, we found it to be still enabled on Ubuntu 17.10 and Debian 9.4, the default operating system for instances on the Google Cloud. 
Moreover, any other function or data can be used instead of vsyscall if the address is known. 
If the address of the \leakGadget is known, it can also be repeated to de-randomize any other function if the execution time of this function can be measured via the network. 
If the attacker knows a memory page at a fixed offset in the kernel, the same attack can also be run on a \netGadget in the kernel to break KASLR.

\section{Evaluation}\label{sec:evaluation}

In this section, we evaluate NetSpectre and the performance of our proof-of-concept implementation.
\Cref{sec:eval:results} provides a qualitative evaluation and \Cref{sec:eval:performance} a quantitative evaluation of our NetSpectre attacks.
For the evaluation we used laptops (Intel Core i5-4200M, i5-6200U, i7-8550U), as well as desktop PCs (Intel Core i7-6700K, i7-8700K), an unspecified Skylake-based Intel Xeon CPU in the Google Cloud Platform, and an ARM Cortex A75.

\subsection{Leakage}\label{sec:eval:results}

To evaluate NetSpectre on the different devices, we constructed a victim program which contains the same \leakGadget and \transmitGadget on all test platforms (\cf~\Cref{sec:overview}).
We leaked known values from the victim to verify that our attack was successful and to determine how many measurements are necessary. 
Except for the cloud setup, all evaluations were done in a local lab environment. 
We used Spectre variant 1 for all evaluations, however, other Spectre variants can be used in the same manner. 

\subsubsection{Desktop and Laptop Computers}\label{sec:eval:desktop}

In contrast to local Spectre attacks, where a single measurement can already be sufficient, NetSpectre attacks require a large number of measurements to distinguish bits with a certain confidence. 
Even on a local network, around \SIx{100000} measurements are required to reduce the noise to a level where a difference between bits can be clearly seen. 
By repeating the attack, the noise is reduced, making it easier to distinguish the bits. 

For our local attack we had a gigabit connection between the victim and the attacker, a typical scenario in local networks but also for network connections of dedicated servers and virtual servers. 
We measured a standard deviation of the network latency of \SI{15.6}{\micro\second}. 
Applying the three-sigma rule~\cite{Pukelsheim1994}, in at least \SI{88.8}{\percent} cases, the latency deviates $\pm$\SI{46.8}{\micro\second} from the average. 
This is nearly 3 orders of magnitude larger than the actual timing difference the attacker wants to measure, explaining the large number of measurements required. 

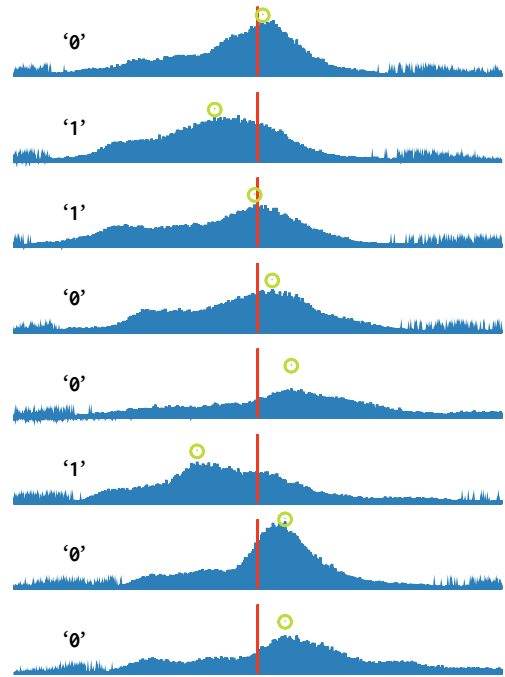
\begin{figureA}[t]{byte_hist}[Leaking the byte `d' (\texttt{01100100} in binary) bit by bit using a NetSpectre attack. The maximum of the histograms (green circle) can be separated using a simple threshold (red line). If the maximum is left of the threshold, the bit is interpreted as `1', otherwise it is interpreted as `0'.]
 \resizebox{0.8\hsize}{!}{
      \begin{tikzpicture}
\begin{axis}[
style={font=\footnotesize},
width=7cm,
scaled y ticks=false,
height=2.5cm,
xmin=14000,
xmax=18000,
legend pos=south east,
axis line style={draw=none},
tick style={draw=none},
ytick=\empty,
xtick=\empty
]
\addplot+[blue,thick,mark=none] table[x=cycle,y=hit,col sep=comma] {data/s_hist_7.csv};
\addplot[green,thick,mark=o] table {
16040 349
16041 350
16042 349
};
\addplot+[red,thick,mark=none] table[x=cycle,y=hit,col sep=comma] {data/s_hist_t.csv};
\node[] at (axis cs: 14500,200) {\textbf{`\texttt{0}'}};


\end{axis}
\end{tikzpicture}
    }
 \resizebox{0.8\hsize}{!}{
      \begin{tikzpicture}
\begin{axis}[
style={font=\footnotesize},
width=7cm,
scaled y ticks=false,
height=2.5cm,
xmin=14000,
xmax=18000,
legend pos=south east,
axis line style={draw=none},
tick style={draw=none},
ytick=\empty,
xtick=\empty
]
\addplot+[blue,thick,mark=none] table[x=cycle,y=hit,col sep=comma] {data/s_hist_6.csv};
\addplot[green,thick,mark=o] table {
15647 299
15648 300
15649 299
};
\addplot+[red,thick,mark=none] table[x=cycle,y=hit,col sep=comma] {data/s_hist_t.csv};
\node[] at (axis cs: 14500,200) {\textbf{`\texttt{1}'}};


\end{axis}
\end{tikzpicture}
    }
 \resizebox{0.8\hsize}{!}{
      \begin{tikzpicture}
\begin{axis}[
style={font=\footnotesize},
width=7cm,
scaled y ticks=false,
height=2.5cm,
xmin=14000,
xmax=18000,
legend pos=south east,
axis line style={draw=none},
tick style={draw=none},
ytick=\empty,
xtick=\empty
]
\addplot+[blue,thick,mark=none] table[x=cycle,y=hit,col sep=comma] {data/s_hist_5.csv};
\addplot[green,thick,mark=o] table {
15974 299
15975 300
15976 299
};
\addplot+[red,thick,mark=none] table[x=cycle,y=hit,col sep=comma] {data/s_hist_t.csv};
\node[] at (axis cs: 14500,200) {\textbf{`\texttt{1}'}};


\end{axis}
\end{tikzpicture}
    }
 \resizebox{0.8\hsize}{!}{
      \begin{tikzpicture}
\begin{axis}[
style={font=\footnotesize},
width=7cm,
scaled y ticks=false,
height=2.5cm,
xmin=14000,
xmax=18000,
legend pos=south east,
axis line style={draw=none},
tick style={draw=none},
ytick=\empty,
xtick=\empty
]
\addplot+[blue,thick,mark=none] table[x=cycle,y=hit,col sep=comma] {data/s_hist_4.csv};
\addplot[green,thick,mark=o] table {
16118 299
16119 300
16120 299
};
\addplot+[red,thick,mark=none] table[x=cycle,y=hit,col sep=comma] {data/s_hist_t.csv};
\node[] at (axis cs: 14500,200) {\textbf{`\texttt{0}'}};


\end{axis}
\end{tikzpicture}
    }
 \resizebox{0.8\hsize}{!}{
      \begin{tikzpicture}
\begin{axis}[
style={font=\footnotesize},
width=7cm,
scaled y ticks=false,
height=2.5cm,
xmin=14000,
xmax=18000,
legend pos=south east,
axis line style={draw=none},
tick style={draw=none},
ytick=\empty,
xtick=\empty
]
\addplot+[blue,thick,mark=none] table[x=cycle,y=hit,col sep=comma] {data/s_hist_3.csv};
\addplot[green,thick,mark=o] table {
16274 299
16275 300
16276 299
};
\addplot+[red,thick,mark=none] table[x=cycle,y=hit,col sep=comma] {data/s_hist_t.csv};
\node[] at (axis cs: 14500,200) {\textbf{`\texttt{0}'}};


\end{axis}
\end{tikzpicture}
    }
 \resizebox{0.8\hsize}{!}{
      \begin{tikzpicture}
\begin{axis}[
style={font=\footnotesize},
width=7cm,
scaled y ticks=false,
height=2.5cm,
xmin=14000,
xmax=18000,
legend pos=south east,
axis line style={draw=none},
tick style={draw=none},
ytick=\empty,
xtick=\empty
]
\addplot+[blue,thick,mark=none] table[x=cycle,y=hit,col sep=comma] {data/s_hist_2.csv};
\addplot[green,thick,mark=o] table {
15503 299
15504 300
15505 299
};
\addplot+[red,thick,mark=none] table[x=cycle,y=hit,col sep=comma] {data/s_hist_t.csv};
\node[] at (axis cs: 14500,200) {\textbf{`\texttt{1}'}};


\end{axis}
\end{tikzpicture}
    }
 \resizebox{0.8\hsize}{!}{
      \begin{tikzpicture}
\begin{axis}[
style={font=\footnotesize},
width=7cm,
scaled y ticks=false,
height=2.5cm,
xmin=14000,
xmax=18000,
legend pos=south east,
axis line style={draw=none},
tick style={draw=none},
ytick=\empty,
xtick=\empty
]
\addplot+[blue,thick,mark=none] table[x=cycle,y=hit,col sep=comma] {data/s_hist_1.csv};
\addplot[green,thick,mark=o] table {
16223 399
16224 400
16225 399
};
\addplot+[red,thick,mark=none] table[x=cycle,y=hit,col sep=comma] {data/s_hist_t.csv};
\node[] at (axis cs: 14500,200) {\textbf{`\texttt{0}'}};


\end{axis}
\end{tikzpicture}
    }
 \resizebox{0.8\hsize}{!}{
      \begin{tikzpicture}
\begin{axis}[
style={font=\footnotesize},
width=7cm,
scaled y ticks=false,
height=2.5cm,
xmin=14000,
xmax=18000,
legend pos=south east,
axis line style={draw=none},
tick style={draw=none},
ytick=\empty,
xtick=\empty
]
\addplot+[blue,thick,mark=none] table[x=cycle,y=hit,col sep=comma] {data/s_hist_0.csv};
\addplot[green,thick,mark=o] table {
16223 299
16224 300
16225 299
};
\addplot+[red,thick,mark=none] table[x=cycle,y=hit,col sep=comma] {data/s_hist_t.csv};

\node[] at (axis cs: 14500,200) {\textbf{`\texttt{0}'}};

\end{axis}
\end{tikzpicture}
    }
\end{figureA}

Our proof-of-concept NetSpectre implementation leaks arbitrary bits from the victim by specifying an out-of-bounds index of a memory bitstream. 
\Cref{fig:byte_hist} shows the leakage of one byte using our proof-of-concept implementation. 
For every bit, we repeated the measurements \SI{1000000} times. 
Although we only use a na\"ive threshold on the maximum of the histograms, we can clearly distinguish `0'-bits from `1'-bits. 
More sophisticated methods, \eg machine learning approaches, might be able to further reduce the number of measurements. 

\subsubsection{ARM Devices}\label{sec:eval:arm}

Also in our evaluation on ARM devices we used a wired network, as the network-latency varies too much in today's wireless connections.
The ARM core we tested turned out to have a significantly higher variance in the network latency. 
We measured a standard deviation of the network latency of \SI{128.5}{\micro\second}. 
Again, with the three-sigma rule, we estimate that at least \SI{88.8}{\percent} of the measurements are within $\pm$\SI{385.5}{\micro\second}.

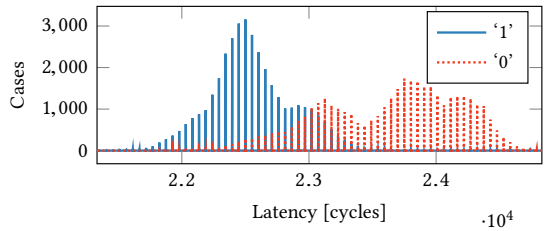
\begin{figureA}[t]{arm_hist}[Histogram of the measurements for a `0'-bit and a `1'-bit on an ARM Cortex A75. Although the times for both cases overlap, they are clearly distinguishable.]
 \resizebox{!}{3cm}{
      \begin{tikzpicture}
\begin{axis}[
style={font=\footnotesize},
xlabel={Latency [cycles]},
ylabel={Cases},
width=7cm,
scaled y ticks=false,
height=3.5cm,
xmin=21333,
xmax=24832,
legend pos=north east
]
\addplot+[blue,thick,mark=none] table[x=cycle,y=hit,col sep=comma] {data/a75_hist_100k.csv};
\addplot+[red,densely dotted,thick,mark=none] table[x=cycle,y=miss,col sep=comma] {data/a75_hist_100k.csv};

\legend{`1',`0'}

\end{axis}
\end{tikzpicture}
    }
\end{figureA}

\Cref{fig:arm_hist} shows two leaked bits---a `0'-bit and a `1'-bit---of an ARM Cortex A75 victim. 
Even with the higher variance in latency, simple thresholding allows separating the maxima of the histograms. 
Hence, the attack also works on ARM devices.

\subsubsection{Cloud Instances}\label{sec:eval:cloud}

For the cloud instance, we tested our proof-of-concept implementation on the Google Cloud Platform. 
We created two virtual machine instances in the same region, one as the attacker, one as the victim. 
For both instances, we used a default Ubuntu 16.04.4 LTS as the operating system. 

The measured standard deviation of the network latency was \SI{52.3}{\micro\second}. 
Thus, we estimate that at least \SI{88.8}{\percent} of the measurements are in a range of $\pm$\SI{156.9}{\micro\second}. 
We verified that we can successfully leak data by running a NetSpectre attack between the two instances.

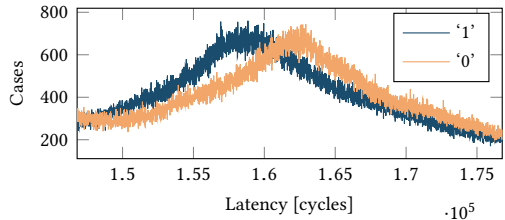
\begin{figureA}[t]{gc_hist}[Histogram of the measurements for a `0'-bit and a `1'-bit on two Google Cloud virtual machines with \SIx{20000000} measurements.]
 \resizebox{!}{3cm}{
      \begin{tikzpicture}
\begin{axis}[
style={font=\footnotesize},
xlabel={Latency [cycles]},
ylabel={Cases},
width=7cm,
scaled y ticks=false,
height=3.5cm,
xmin=146794,
xmax=176782,
legend pos=north east
]
\addplot+[blue!60!black,mark=none] table[x=cycle,y=hit,col sep=comma] {data/small_gc_20mio.csv};
\addplot+[red!40!yellow,mark=none] table[x=cycle,y=miss,col sep=comma] {data/small_gc_20mio.csv};

\legend{`1',`0'}

\end{axis}
\end{tikzpicture}
    }
\end{figureA}

To adapt for the higher variance in network latency, we increased the number of measurements by a factor of \SIx{20}, \ie every bit was measured \SI{20000000} times. 
\Cref{fig:gc_hist} shows a (smoothed) histogram for both a `0'-bit and a `1'-bit on the Google Cloud instances. 
Although there is still noise visible, it is possible to distinguish the bits and thus leak arbitrary bits from the victim cloud instance.

\subsection{NetSpectre Performance}\label{sec:eval:performance}
To evaluate the performance of NetSpectre, we leaked known values from a target device.
This allows us to not only determine how fast an attacker can leak memory, but also to determine the bit-error rate, \ie how many bit errors to expect.

\subsubsection{Local Network}
Attacks on the local network achieve the best performance, as the variance in network latency is significantly smaller than the variance over the internet (\cf \Cref{sec:eval:cloud}). 
In our lab setup, we repeat the measurement \SI{1000000} times per bit to be able to reliably leak bytes from the victim. 
On average, leaking one byte takes \SI{30}{\minute}, which amounts to approximately \SI{4}{\minute} per bit.
Using the AVX covert channel instead of the cache reduces the required time to leak an entire byte to only \SI{8}{\minute}. 

To break ASLR, we require the cache covert channel. 
On average, this allows breaking the randomization remotely within \SI{2}{\hour}.

We used \texttt{stress -i 1 -d 1} for the experiments, to simulate a realistic environment. 
Although we would have expected our attack to work best on a completely idle server, we did not see any negative effects from the moderate server loads.
In fact, they even slightly improved the attack performance. 
One reason for this is that a higher server load incurs a higher number of memory and cache accesses~\cite{Aciicmez2007d} and thus facilitates the cache thrashing (\cf \cref{sec:covert_channels}), which is the performance bottle neck of our attack. 
Another reason is that a higher server load might exhaust execution ports required to calculate the bounds check in the leak gadget, thus increasing the chance that the CPU has to execute the condition speculatively. 

Our NetSpectre attack in local networks is comparably slow.
However, in particular specialized malware attacks, \eg APTs, are often active over several months in local networks.
Over such a time frame, the attacker can indeed leak all data of interest from a target system on the same network.

\subsubsection{Cloud Network}

We evaluated the performance in the cloud using two virtual machines instances on the Google Cloud.
These virtual machines have a fast network connection. 
We configured the two instances to each use 2 virtual CPUs, which enabled a \SI{4}{\giga bit/\second} connection~\cite{GC2018throughput}. 
In this setup, we repeat the measurement \SI{20000000} times per bit to get an error-free leakage of bytes. 
On average, leaking one byte takes \SI{8}{\hour} for the cache covert channel, and \SI{3}{\hour} for the AVX covert channel. 

While this is comparably slow, it shows that remote Spectre attacks are feasible between independent instances in the public cloud.
In particular, APTs typically run for several weeks or months.
Such an extended time frame is clearly sufficient to leak sensitive data, such as encryption keys or passwords, using the NetSpectre attack in a cloud environment.

\section{Challenges of Mitigating Spectre}\label{sec:countermeasures}

In this section, we discuss limitations of state-of-the-art countermeasures against Spectre, and how they do not fully prevent NetSpectre attacks. 
Furthermore, we discuss how NetSpectre attacks can be prevented on the network layer. 
Finally, we outline challenges for future research on Spectre attacks as well as Spectre mitigations.

\subsection{State-of-the-art Spectre Countermeasures}
Due to the different origins, Spectre variant 1 and variant 2 are mitigated using separate countermeasures.
Intel released microcode updates to prevent the cross-process and cross-privilege mistraining of indirect branches typical for Spectre variant 2 attacks. 
There are no microcode updates to prevent mistraining of direct branches, since this is easy to do in-place, \ie in the same privilege level and the same process context. 
For Spectre variant 1 attacks, a series of pure software countermeasures have been proposed.

Intel and AMD recommend using the \texttt{lfence} instruction as a speculation barrier~\cite{IntelSpecAnalysis,AMDSpecAnalysis}. 
This instruction has to be inserted after security-critical bounds check to stop the speculative execution. 
However, adding this to every bounds check has a significant performance overhead~\cite{IntelSpecAnalysis}.

Moreover, our experiments showed that \texttt{lfence}s do stop the speculative execution but not speculative code fetches and other microarchitectural behaviors that occur pre-execution, such as powering up of the AVX functional units, instruction cache fills, and TLB fills. 
According to our experiments, \texttt{lfence}s do work against traditional Spectre gadgets but not against all Spectre gadgets we use in this paper, \cf \Cref{lst:avx2_leak} and \Cref{lst:aslr_gadget}, which can already leak information through microarchitectural behaviors that occur pre-execution. 
However, we believe there are ways to use \texttt{lfence}s in a way that mitigates the leakage.

Microsoft implements an automatic detection of vulnerable code paths, \ie Spectre gadgets, in its compiler to limit the speculation barrier to these gadgets~\cite{Microsoft2018Spectre}. 
However, \citeA{Kocher2018mitigations} showed that the automated analysis misses many gadgets. 
As Microsoft only uses a blacklist for known gadgets~\cite{Kocher2018mitigations}, many gadgets, in particular gadgets which are not typical (\eg gadgets to break ASLR), are not automatically safeguarded by the compiler. 

In the Linux kernel, exploitable gadgets are identified manually and with the help of static code analyzers~\cite{LWN2018spectrecoccinelle}. 
Similarly to the compiler-based approach, this requires a complete understanding of which code snippets are exploitable. 

Finally, until now it was widely overlooked that indirect branch mistraining (Spectre variant 2) is also possible in-place.
However, the attack possibilities are much more constrained with in-place mistraining.

\subsection{Network-layer Countermeasures}
As NetSpectre is a network-based attack, it cannot only be prevented by mitigating Spectre but also through countermeasures on the network layer. 
A trivial NetSpectre attack can easily be detected by a DDoS protection, as multiple thousand identical packets are sent from the same source. 
However, an attacker can choose any trade-off between packets per second and leaked bits per second. 
Thus, the speed at which bits are leaked can simply be reduced below the threshold that the DDoS monitoring can detect. 
This is true for any monitoring which tries to detect ongoing attacks, \eg intrusion detection systems. 
Although the attack is theoretically not prevented, at some point the attack becomes infeasible, as the time required to leak a bit increases drastically. 

Another method to mitigate NetSpectre is to add artificial noise to the network latency. 
As the number of measurements depends on the variance in network latency, additional noise requires an attacker to perform more measurements. 
Thus, if the variance in network latency is high enough, NetSpectre attacks become infeasible due to the large number of measurements required. 

Both approaches may mitigate NetSpectre attacks in practice.
However, as attackers can adapt and improve attacks, it is not safe to assume that noise levels and monitoring thresholds chosen now will still be valid in the near future.

\subsection{Future Research Challenges}
As discussed in the previous sections, Spectre is far from being a solved case.
The currently proposed mitigations merely fix symptoms without directly addressing the root cause, the imbalanced trade-off between performance and security that led to the speculative execution we currently have. 
We identified 5 challenges (C1 to C5) for future work on Spectre attacks and mitigations. 

\paragraph{\textbf{C1}: Gadgets are more versatile than anticipated.}
In particular the gadgets we use to break ASLR have not been considered dangerous so far.
Also the AVX-based gadgets we use were not considered so far.
Gadgets may also consist of many small code pieces that pass on secret values until at a later point the secret value is leaked to the attacker.
Since the building block of Spectre that exfiltrates the information to the attacker is a covert channel, it appears the underlying problem of identifying all gadgets may be reduced to the problem of identifying all covert channels.
Currently, we have no technique to identify all covert channels in a system.

\paragraph{\textbf{C2}: Automatically safeguarding all gadgets is not trivial.}
For Spectre variant 1 the proposed solution is to use speculation barriers~\cite{IntelSpecAnalysis,AMDSpecAnalysis}. 
As we cannot expect every developer to identify vulnerable gadgets and correctly fix them, state-of-the-art solutions try to automatically detect vulnerable gadgets and fix them at compile time~\cite{Microsoft2018Spectre}. 
At the moment it is not clear whether static code analysis is sufficient to detect all vulnerable gadgets, especially if they are scattered across functions. 
In such complex scenarios, dynamic analysis might lead to better results. 
However, dynamic analysis naturally suffers from incompleteness, as certain parts of the program may not be reached in the dynamic analysis.
Furthermore, it might be possible that the compiler produces Spectre gadgets which are not visible in the source code, as it can happen with \eg double-fetch bugs~\cite{Schwarz2018DF}.
This would hardly be detected upfront and completely undermine the security measures taken.

\paragraph{\textbf{C3}: Blacklisting is inherently incomplete.}
Current approaches rely on blacklists to automatically patch exploitable gadgets~\cite{Microsoft2018Spectre}. 
However, this implies that we understand exactly which code fragments are exploitable and which are not. 
As this paper shows, gadgets can look different than anticipated, showing the incompleteness of the blacklist approach.
Inverting the logic might be a better direction, \ie using whitelists of (provably) unexploitable gadgets instead of blacklists. 
However, this would require a substantial amount of research on proving non-exploitability of code fragments.

\paragraph{\textbf{C4}: Cross-process and cross-privilege-level mistraining is easier to solve than in-place mistraining.}
Current countermeasures mainly aim at preventing Spectre attacks across process boundaries and there especially across privilege levels~\cite{Intel2018spectrefix,IntelSpecAnalysis,ARMSpecAnalysis}. 
However, as shown in this paper, such countermeasures are ineffective if the mistraining happens in-place inside the same process. 
This method is not only applicable to Spectre variant 1, but also to Spectre variant 2~\cite{Gruss2018Bluehat}. 
Retpoline~\cite{Turner2018retpoline} is currently the only mitigation that protects against these Spectre variant 2 attacks by effectively stopping any further speculation by the processor.
However, Retpoline is not a perfect solution, as it incurs significant performance overheads and adds another side channel~\cite{Fogh2018retpoline}. 

If an attacker can only poison branches with valid branch targets inside the same process, \ie all microcode updates applied, Retpoline can be replaced by a more simple construct we propose.
We propose to insert speculation barriers at every possible call target.
This is a much clearer structure than with Retpolines.
Thus, every misspeculated indirect call immediately aborts before actually executing code. 
For direct calls, the compiler can jump just beyond the speculation barrier to have less performance impact.
Still, the overall performance impact of this solution, just like Retpoline, would be significant.
It remains unclear whether Spectre attacks within the same process can be fully prevented without high performance overheads and without introducing new problems. 

\paragraph{\textbf{C5}: Security mechanisms may have unwanted side effects.}
The Retpoline patch basically hides the target of indirect calls from the CPU by fiddling with the return values on the stack. 
However, this leads to side effects with other security mechanisms, as a Retpoline behaves similarly to an exploit changing the control flow. 
Especially security mechanisms such as control-flow integrity have to be adapted~\cite{GCC2018retpolinecfi} to not falsely detect Retpolines as attacks. 
Still, the question arises how Spectre mitigations interact with other CFI implementations, especially in hardware, as well as other security mechanisms and whether we have to accept trade-offs when combining security mechanisms. 
In general, we need to investigate which security mechanisms could have detrimental side effects that outweigh the gains in security.

\section{Conclusion}\label{sec:conclusion}
In this paper, we presented NetSpectre, the first remote Spectre variant 1 attack. 
We demonstrated the first access-driven remote \EvictReload cache attack over network, with a performance of 15 bits per hour.
We also demonstrated the first Spectre attack which does not use a cache covert channel.
In particular, in a remote Spectre attack, our novel high-performance AVX-based covert channel performs significantly better than the remote cache covert channel.
Our NetSpectre attack in combination with the AVX-based covert channel leaks 60 bits per hour from the target system.
We verified NetSpectre in local networks as well as in the Google cloud.

NetSpectre marks a paradigm shift for Spectre attacks, from local attacks to remote attacks.
With our NetSpectre attacks, a much wider range and larger number of devices are exposed to Spectre attacks.
Spectre attacks now must also be considered on devices which do not run any potentially attacker-controlled code at all.
We demonstrate that in a remote attack, NetSpectre can be used to defeat address-space layout randomization on the remote system.
As we discussed in this paper, there are a series of open challenges for future research on Spectre attacks and Spectre mitigations.


\section*{Acknowledgments}
We would like to thank Anders Fogh, Halvar Flake, Jann Horn, Stefan Mangard, and Matt Miller for feedback on an early draft.

This work has been supported by the European Research Council (ERC) under the European Union's Horizon 2020 research and innovation programme (grant agreement No 681402).

%
\bibliographystyle{ACM-Reference-Format}
\bibliography{main}  


\begin{thebibliography}{00}


\ifx \showCODEN    \undefined \def \showCODEN     #1{\unskip}     \fi
\ifx \showDOI      \undefined \def \showDOI       #1{#1}\fi
\ifx \showISBNx    \undefined \def \showISBNx     #1{\unskip}     \fi
\ifx \showISBNxiii \undefined \def \showISBNxiii  #1{\unskip}     \fi
\ifx \showISSN     \undefined \def \showISSN      #1{\unskip}     \fi
\ifx \showLCCN     \undefined \def \showLCCN      #1{\unskip}     \fi
\ifx \shownote     \undefined \def \shownote      #1{#1}          \fi
\ifx \showarticletitle \undefined \def \showarticletitle #1{#1}   \fi
\ifx \showURL      \undefined \def \showURL       {\relax}        \fi
\providecommand\bibfield[2]{#2}
\providecommand\bibinfo[2]{#2}
\providecommand\natexlab[1]{#1}
\providecommand\showeprint[2][]{arXiv:#2}

\bibitem[\protect\citeauthoryear{Ac\i{}i\c{c}mez, Schindler, and
  Ko\c{c}}{Ac\i{}i\c{c}mez et~al\mbox{.}}{2007}]%
        {Aciicmez2007d}
\bibfield{author}{\bibinfo{person}{Onur Ac\i{}i\c{c}mez},
  \bibinfo{person}{Werner Schindler}, {and} \bibinfo{person}{\c{C}etin~Kaya
  Ko\c{c}}.} \bibinfo{year}{2007}\natexlab{}.
\newblock \showarticletitle{{Cache Based Remote Timing Attack on the AES}}. In
  \bibinfo{booktitle}{{\em CT-RSA 2007}}.
\newblock


\bibitem[\protect\citeauthoryear{{Advanced Micro Devices, Inc.}}{{Advanced
  Micro Devices, Inc.}}{2002}]%
        {ARMGuide2002realview}
\bibfield{author}{\bibinfo{person}{{Advanced Micro Devices, Inc.}}}
  \bibinfo{year}{2002}\natexlab{}.
\newblock \bibinfo{title}{RealView{\textregistered} Compilation Tools}.
\newblock   (\bibinfo{year}{2002}).
\newblock


\bibitem[\protect\citeauthoryear{{Advanced Micro Devices, Inc.}}{{Advanced
  Micro Devices, Inc.}}{2017}]%
        {AMD64_manual}
\bibfield{author}{\bibinfo{person}{{Advanced Micro Devices, Inc.}}}
  \bibinfo{year}{2017}\natexlab{}.
\newblock \bibinfo{title}{{AMD64 Architecture Programmer's Manual}}.
\newblock   (\bibinfo{year}{2017}).
\newblock


\bibitem[\protect\citeauthoryear{{Advanced Micro Devices, Inc.}}{{Advanced
  Micro Devices, Inc.}}{2018}]%
        {AMDSpecAnalysis}
\bibfield{author}{\bibinfo{person}{{Advanced Micro Devices, Inc.}}}
  \bibinfo{year}{2018}\natexlab{}.
\newblock \bibinfo{title}{Software Techniques for Managing Speculation on {AMD}
  Processors}.
\newblock
  \bibinfo{howpublished}{\url{http://developer.amd.com/wordpress/media/2013/12/Managing-Speculation-on-AMD-Processors.pdf}}.
    (\bibinfo{year}{2018}).
\newblock


\bibitem[\protect\citeauthoryear{Aly and ElGayyar}{Aly and ElGayyar}{2013}]%
        {Aly2013attacking}
\bibfield{author}{\bibinfo{person}{Hassan Aly} {and} \bibinfo{person}{Mohammed
  ElGayyar}.} \bibinfo{year}{2013}\natexlab{}.
\newblock \showarticletitle{Attacking aes using bernstein's attack on modern
  processors}. In \bibinfo{booktitle}{{\em International Conference on
  Cryptology in Africa}}.
\newblock


\bibitem[\protect\citeauthoryear{{ARM}}{{ARM}}{2018}]%
        {ARMSpecAnalysis}
\bibfield{author}{\bibinfo{person}{{ARM}}.} \bibinfo{year}{2018}\natexlab{}.
\newblock \bibinfo{title}{Vulnerability of Speculative Processors to Cache
  Timing Side-Channel Mechanism}.
\newblock
  \bibinfo{howpublished}{\url{https://developer.arm.com/support/security-update}}.
    (\bibinfo{year}{2018}).
\newblock


\bibitem[\protect\citeauthoryear{{ARM Limited}}{{ARM Limited}}{2012}]%
        {arm_arch_manualv7}
\bibfield{author}{\bibinfo{person}{{ARM Limited}}.}
  \bibinfo{year}{2012}\natexlab{}.
\newblock \bibinfo{booktitle}{{\em ARM Architecture Reference Manual. ARMv7-A
  and ARMv7-R edition}}.
\newblock \bibinfo{publisher}{{ARM Limited}}.
\newblock


\bibitem[\protect\citeauthoryear{{ARM Limited}}{{ARM Limited}}{2013}]%
        {arm_arch_manualv8}
\bibfield{author}{\bibinfo{person}{{ARM Limited}}.}
  \bibinfo{year}{2013}\natexlab{}.
\newblock \bibinfo{booktitle}{{\em {ARM Architecture Reference Manual ARMv8}}}.
\newblock \bibinfo{publisher}{{ARM Limited}}.
\newblock


\bibitem[\protect\citeauthoryear{Bencs{\'a}th, P{\'e}k, Butty{\'a}n, and
  F{\'e}legyh{\'a}zi}{Bencs{\'a}th et~al\mbox{.}}{2012}]%
        {Bencsath2012duqu}
\bibfield{author}{\bibinfo{person}{Boldizs{\'a}r Bencs{\'a}th},
  \bibinfo{person}{G{\'a}bor P{\'e}k}, \bibinfo{person}{Levente Butty{\'a}n},
  {and} \bibinfo{person}{M{\'a}rk F{\'e}legyh{\'a}zi}.}
  \bibinfo{year}{2012}\natexlab{}.
\newblock \showarticletitle{Duqu: Analysis, detection, and lessons learned}. In
  \bibinfo{booktitle}{{\em ACM European Workshop on System Security
  (EuroSec)}}, Vol.~\bibinfo{volume}{2012}.
\newblock


\bibitem[\protect\citeauthoryear{Benger, van~de Pol, Smart, and Yarom}{Benger
  et~al\mbox{.}}{2014}]%
        {Benger2014}
\bibfield{author}{\bibinfo{person}{Naomi Benger}, \bibinfo{person}{Joop van~de
  Pol}, \bibinfo{person}{Nigel~P Smart}, {and} \bibinfo{person}{Yuval Yarom}.}
  \bibinfo{year}{2014}\natexlab{}.
\newblock \showarticletitle{{\enquote{Ooh Aah... Just a Little Bit}: A small
  amount of side channel can go a long way}}. In \bibinfo{booktitle}{{\em
  CHES'14}}.
\newblock


\bibitem[\protect\citeauthoryear{Bernstein}{Bernstein}{2005}]%
        {Bernstein2005}
\bibfield{author}{\bibinfo{person}{Daniel~J. Bernstein}.}
  \bibinfo{year}{2005}\natexlab{}.
\newblock \bibinfo{title}{{Cache-Timing Attacks on AES}}.
\newblock   (\bibinfo{year}{2005}).
\newblock
\showURL{%
\url{http://cr.yp.to/antiforgery/cachetiming-20050414.pdf}}


\bibitem[\protect\citeauthoryear{Bhattacharya, Maurice, Bhasin, and
  Mukhopadhyay}{Bhattacharya et~al\mbox{.}}{2017}]%
        {Bhattacharya2017perf}
\bibfield{author}{\bibinfo{person}{Sarani Bhattacharya},
  \bibinfo{person}{Cl\'ementine Maurice}, \bibinfo{person}{Shivam Bhasin},
  {and} \bibinfo{person}{Debdeep Mukhopadhyay}.}
  \bibinfo{year}{2017}\natexlab{}.
\newblock \bibinfo{title}{Template Attack on Blinded Scalar Multiplication with
  Asynchronous perf-ioctl Calls}.
\newblock \bibinfo{howpublished}{Cryptology ePrint Archive, Report 2017/968}.
  (\bibinfo{year}{2017}).
\newblock


\bibitem[\protect\citeauthoryear{Bulygin}{Bulygin}{2008}]%
        {Bulygin2008cpu}
\bibfield{author}{\bibinfo{person}{Yuriy Bulygin}.}
  \bibinfo{year}{2008}\natexlab{}.
\newblock \showarticletitle{Cpu side-channels vs. virtualization malware: The
  good, the bad, or the ugly}.
\newblock \bibinfo{journal}{{\em Proceedings of the ToorCon\/}}
  (\bibinfo{year}{2008}).
\newblock


\bibitem[\protect\citeauthoryear{Charneski}{Charneski}{2015}]%
        {Charneski2015}
\bibfield{author}{\bibinfo{person}{Andrew Charneski}.}
  \bibinfo{year}{2015}\natexlab{}.
\newblock \bibinfo{title}{Modeling Network Latency}.
\newblock   (\bibinfo{year}{2015}).
\newblock
\showURL{%
\url{https://blog.simiacryptus.com/2015/10/modeling-network-latency.html}}


\bibitem[\protect\citeauthoryear{Chen, Chen, Xiao, Zhang, Lin, and Lai}{Chen
  et~al\mbox{.}}{2018}]%
        {Chen2018SGXpectre}
\bibfield{author}{\bibinfo{person}{Guoxing Chen}, \bibinfo{person}{Sanchuan
  Chen}, \bibinfo{person}{Yuan Xiao}, \bibinfo{person}{Yinqian Zhang},
  \bibinfo{person}{Zhiqiang Lin}, {and} \bibinfo{person}{Ten~H Lai}.}
  \bibinfo{year}{2018}\natexlab{}.
\newblock \showarticletitle{SGXPECTRE Attacks: Leaking Enclave Secrets via
  Speculative Execution}.
\newblock \bibinfo{journal}{{\em arXiv:1802.09085\/}} (\bibinfo{year}{2018}).
\newblock


\bibitem[\protect\citeauthoryear{Cock, Ge, Murray, and Heiser}{Cock
  et~al\mbox{.}}{2014}]%
        {Cock2014timing}
\bibfield{author}{\bibinfo{person}{David Cock}, \bibinfo{person}{Qian Ge},
  \bibinfo{person}{Toby Murray}, {and} \bibinfo{person}{Gernot Heiser}.}
  \bibinfo{year}{2014}\natexlab{}.
\newblock \showarticletitle{The last mile: An empirical study of timing
  channels on seL4}. In \bibinfo{booktitle}{{\em Proceedings of the 2014 ACM
  SIGSAC Conference on Computer and Communications Security}}.
\newblock


\bibitem[\protect\citeauthoryear{Corbet}{Corbet}{2011}]%
        {LWN2011vsyscall}
\bibfield{author}{\bibinfo{person}{Jonathan Corbet}.}
  \bibinfo{year}{2011}\natexlab{}.
\newblock \bibinfo{title}{On vsyscalls and the vDSO}.
\newblock   (\bibinfo{year}{2011}).
\newblock
\showURL{%
\url{https://lwn.net/Articles/446528/}}


\bibitem[\protect\citeauthoryear{Evtyushkin, Ponomarev, and
  Abu-Ghazaleh}{Evtyushkin et~al\mbox{.}}{2016}]%
        {Evtyushkin2016ASLR}
\bibfield{author}{\bibinfo{person}{Dmitry Evtyushkin}, \bibinfo{person}{Dmitry
  Ponomarev}, {and} \bibinfo{person}{Nael Abu-Ghazaleh}.}
  \bibinfo{year}{2016}\natexlab{}.
\newblock \showarticletitle{Jump over ASLR: Attacking branch predictors to
  bypass ASLR}. In \bibinfo{booktitle}{{\em International Symposium on
  Microarchitecture (MICRO)}}.
\newblock


\bibitem[\protect\citeauthoryear{Fog}{Fog}{2015}]%
        {Fog2015AVXb}
\bibfield{author}{\bibinfo{person}{Agner Fog}.}
  \bibinfo{year}{2015}\natexlab{}.
\newblock \bibinfo{title}{Test results for Broadwell and Skylake}.
\newblock   (\bibinfo{year}{2015}).
\newblock
\showURL{%
\url{http://www.agner.org/optimize/blog/read.php?i=415#415}}


\bibitem[\protect\citeauthoryear{Fog}{Fog}{2016}]%
        {Fog2016}
\bibfield{author}{\bibinfo{person}{Agner Fog}.}
  \bibinfo{year}{2016}\natexlab{}.
\newblock \bibinfo{booktitle}{{\em The microarchitecture of {Intel}, {AMD} and
  {VIA} {CPU}s: An optimization guide for assembly programmers and compiler
  makers}}.
\newblock
\showURL{%
\url{http://www.agner.org/optimize/microarchitecture.pdf}}


\bibitem[\protect\citeauthoryear{Fogh}{Fogh}{2018}]%
        {Fogh2018retpoline}
\bibfield{author}{\bibinfo{person}{Anders Fogh}.}
  \bibinfo{year}{2018}\natexlab{}.
\newblock \bibinfo{title}{{In debt to Retpoline}}.
\newblock   (\bibinfo{year}{2018}).
\newblock
\showURL{%
\url{https://cyber.wtf/2018/02/13/in-debt-to-retpoline/}}


\bibitem[\protect\citeauthoryear{Ge, Yarom, Cock, and Heiser}{Ge
  et~al\mbox{.}}{2016}]%
        {Ge2016}
\bibfield{author}{\bibinfo{person}{Qian Ge}, \bibinfo{person}{Yuval Yarom},
  \bibinfo{person}{David Cock}, {and} \bibinfo{person}{Gernot Heiser}.}
  \bibinfo{year}{2016}\natexlab{}.
\newblock \showarticletitle{{A Survey of Microarchitectural Timing Attacks and
  Countermeasures on Contemporary Hardware}}.
\newblock \bibinfo{journal}{{\em Journal of Cryptographic Engineering\/}}
  (\bibinfo{year}{2016}).
\newblock


\bibitem[\protect\citeauthoryear{Google}{Google}{2018}]%
        {GC2018throughput}
\bibfield{author}{\bibinfo{person}{Google}.} \bibinfo{year}{2018}\natexlab{}.
\newblock \bibinfo{title}{Egress throughput caps}.
\newblock   (\bibinfo{year}{2018}).
\newblock
\showURL{%
\url{https://cloud.google.com/compute/docs/networks-and-firewalls#egress_throughput_caps}}


\bibitem[\protect\citeauthoryear{Goonatilake and Bachnak}{Goonatilake and
  Bachnak}{2012}]%
        {Goonatilake2012Latency}
\bibfield{author}{\bibinfo{person}{Rohitha Goonatilake} {and}
  \bibinfo{person}{Rafic~A Bachnak}.} \bibinfo{year}{2012}\natexlab{}.
\newblock \showarticletitle{Modeling latency in a network distribution}.
\newblock \bibinfo{journal}{{\em Network and Communication Technologies\/}}
  \bibinfo{volume}{1}, \bibinfo{number}{2} (\bibinfo{year}{2012}),
  \bibinfo{pages}{1}.
\newblock


\bibitem[\protect\citeauthoryear{Gras, Razavi, Bosman, Bos, and Giuffrida}{Gras
  et~al\mbox{.}}{2017}]%
        {Gras2017AnC}
\bibfield{author}{\bibinfo{person}{Ben Gras}, \bibinfo{person}{Kaveh Razavi},
  \bibinfo{person}{Erik Bosman}, \bibinfo{person}{Herbert Bos}, {and}
  \bibinfo{person}{Cristiano Giuffrida}.} \bibinfo{year}{2017}\natexlab{}.
\newblock \showarticletitle{{ASLR} on the {Line}: {Practical} {Cache} {Attacks}
  on the {MMU}}. In \bibinfo{booktitle}{{\em {NDSS}}}.
\newblock


\bibitem[\protect\citeauthoryear{Gruss, Lipp, and Schwarz}{Gruss
  et~al\mbox{.}}{2018}]%
        {Gruss2018Bluehat}
\bibfield{author}{\bibinfo{person}{Daniel Gruss}, \bibinfo{person}{Moritz
  Lipp}, {and} \bibinfo{person}{Michael Schwarz}.}
  \bibinfo{year}{2018}\natexlab{}.
\newblock \showarticletitle{{Beyond Belief: The Case of Spectre and Meltdown}}.
\newblock \bibinfo{journal}{{\em Bluehat IL\/}} (\bibinfo{date}{Jan.}
  \bibinfo{year}{2018}).
\newblock
\showURL{%
\url{http://www.bluehatil.com/files/Beyond%20Belief%20-%20The%20Case%20of%20Spectre%20and%20Meltdown.pdf}}


\bibitem[\protect\citeauthoryear{Gruss, Lipp, Schwarz, Fellner, Maurice, and
  Mangard}{Gruss et~al\mbox{.}}{2017}]%
        {Gruss2017KASLR}
\bibfield{author}{\bibinfo{person}{Daniel Gruss}, \bibinfo{person}{Moritz
  Lipp}, \bibinfo{person}{Michael Schwarz}, \bibinfo{person}{Richard Fellner},
  \bibinfo{person}{Cl{\'e}mentine Maurice}, {and} \bibinfo{person}{Stefan
  Mangard}.} \bibinfo{year}{2017}\natexlab{}.
\newblock \showarticletitle{{KASLR is Dead: Long Live KASLR}}. In
  \bibinfo{booktitle}{{\em {ESSoS}}}.
\newblock


\bibitem[\protect\citeauthoryear{Gruss, Maurice, Fogh, Lipp, and Mangard}{Gruss
  et~al\mbox{.}}{2016a}]%
        {Gruss2016Prefetch}
\bibfield{author}{\bibinfo{person}{Daniel Gruss}, \bibinfo{person}{Clémentine
  Maurice}, \bibinfo{person}{Anders Fogh}, \bibinfo{person}{Moritz Lipp}, {and}
  \bibinfo{person}{Stefan Mangard}.} \bibinfo{year}{2016}\natexlab{a}.
\newblock \showarticletitle{{Prefetch Side-Channel Attacks: Bypassing SMAP and
  Kernel ASLR}}. In \bibinfo{booktitle}{{\em CCS}}.
\newblock


\bibitem[\protect\citeauthoryear{Gruss, Maurice, Wagner, and Mangard}{Gruss
  et~al\mbox{.}}{2016b}]%
        {Gruss2016Flush}
\bibfield{author}{\bibinfo{person}{Daniel Gruss},
  \bibinfo{person}{Cl\'{e}mentine Maurice}, \bibinfo{person}{Klaus Wagner},
  {and} \bibinfo{person}{Stefan Mangard}.} \bibinfo{year}{2016}\natexlab{b}.
\newblock \showarticletitle{{Flush+Flush: A Fast and Stealthy Cache Attack}}.
  In \bibinfo{booktitle}{{\em DIMVA}}.
\newblock


\bibitem[\protect\citeauthoryear{Gruss, Spreitzer, and Mangard}{Gruss
  et~al\mbox{.}}{2015}]%
        {Gruss2015Template}
\bibfield{author}{\bibinfo{person}{Daniel Gruss}, \bibinfo{person}{Raphael
  Spreitzer}, {and} \bibinfo{person}{Stefan Mangard}.}
  \bibinfo{year}{2015}\natexlab{}.
\newblock \showarticletitle{{Cache Template Attacks: Automating Attacks on
  Inclusive Last-Level Caches}}. In \bibinfo{booktitle}{{\em USENIX Security
  Symposium}}.
\newblock


\bibitem[\protect\citeauthoryear{Gullasch, Bangerter, and Krenn}{Gullasch
  et~al\mbox{.}}{2011}]%
        {Gullasch2011}
\bibfield{author}{\bibinfo{person}{David Gullasch}, \bibinfo{person}{Endre
  Bangerter}, {and} \bibinfo{person}{Stephan Krenn}.}
  \bibinfo{year}{2011}\natexlab{}.
\newblock \showarticletitle{{Cache Games -- Bringing Access-Based Cache Attacks
  on AES to Practice}}. In \bibinfo{booktitle}{{\em S\&P}}.
\newblock


\bibitem[\protect\citeauthoryear{Guri, Kedma, Kachlon, and Elovici}{Guri
  et~al\mbox{.}}{2014}]%
        {Guri2014airhopper}
\bibfield{author}{\bibinfo{person}{Mordechai Guri}, \bibinfo{person}{Gabi
  Kedma}, \bibinfo{person}{Assaf Kachlon}, {and} \bibinfo{person}{Yuval
  Elovici}.} \bibinfo{year}{2014}\natexlab{}.
\newblock \showarticletitle{AirHopper: Bridging the air-gap between isolated
  networks and mobile phones using radio frequencies}. In
  \bibinfo{booktitle}{{\em 9th International Conference on Malicious and
  Unwanted: Software The Americas (MALWARE)}}.
\newblock


\bibitem[\protect\citeauthoryear{Guri, Monitz, Mirski, and Elovici}{Guri
  et~al\mbox{.}}{2015}]%
        {Guri2015bitwhisper}
\bibfield{author}{\bibinfo{person}{Mordechai Guri}, \bibinfo{person}{Matan
  Monitz}, \bibinfo{person}{Yisroel Mirski}, {and} \bibinfo{person}{Yuval
  Elovici}.} \bibinfo{year}{2015}\natexlab{}.
\newblock \showarticletitle{Bitwhisper: Covert signaling channel between
  air-gapped computers using thermal manipulations}. In
  \bibinfo{booktitle}{{\em IEEE 28th Computer Security Foundations Symposium
  (CSF)}}.
\newblock


\bibitem[\protect\citeauthoryear{Hopper, Vasserman, and Chan-Tin}{Hopper
  et~al\mbox{.}}{2010}]%
        {Hopper2010Tor}
\bibfield{author}{\bibinfo{person}{Nicholas Hopper}, \bibinfo{person}{Eugene~Y
  Vasserman}, {and} \bibinfo{person}{Eric Chan-Tin}.}
  \bibinfo{year}{2010}\natexlab{}.
\newblock \showarticletitle{How much anonymity does network latency leak?}
\newblock \bibinfo{journal}{{\em ACM Transactions on Information and System
  Security (TISSEC)\/}} (\bibinfo{year}{2010}).
\newblock


\bibitem[\protect\citeauthoryear{Hund, Willems, and Holz}{Hund
  et~al\mbox{.}}{2013}]%
        {Hund2013}
\bibfield{author}{\bibinfo{person}{Ralf Hund}, \bibinfo{person}{Carsten
  Willems}, {and} \bibinfo{person}{Thorsten Holz}.}
  \bibinfo{year}{2013}\natexlab{}.
\newblock \showarticletitle{{Practical Timing Side Channel Attacks against
  Kernel Space ASLR}}. In \bibinfo{booktitle}{{\em {S\&P}}}.
\newblock


\bibitem[\protect\citeauthoryear{Intel}{Intel}{2014}]%
        {Intel_vol2}
\bibfield{author}{\bibinfo{person}{Intel}.} \bibinfo{year}{2014}\natexlab{}.
\newblock \showarticletitle{{Intel{$\textregistered$} 64 and IA-32
  Architectures Software Developer{$\textquoteright$}s Manual Volume 2 (2A, 2B
  \& 2C): Instruction Set Reference, A-Z}}.
\newblock   \bibinfo{volume}{253665} (\bibinfo{year}{2014}).
\newblock


\bibitem[\protect\citeauthoryear{Intel}{Intel}{2016a}]%
        {Intel_vol1}
\bibfield{author}{\bibinfo{person}{Intel}.} \bibinfo{year}{2016}\natexlab{a}.
\newblock \showarticletitle{{Intel{$\textregistered$} 64 and IA-32
  Architectures Software Developer{$\textquoteright$}s Manual, Volume 1: Basic
  Architecture}}.
\newblock   \bibinfo{volume}{253665} (\bibinfo{year}{2016}).
\newblock


\bibitem[\protect\citeauthoryear{Intel}{Intel}{2016b}]%
        {Intel_vol3}
\bibfield{author}{\bibinfo{person}{Intel}.} \bibinfo{year}{2016}\natexlab{b}.
\newblock \showarticletitle{{Intel{$\textregistered$} 64 and IA-32
  Architectures Software Developer{$\textquoteright$}s Manual, Volume 3 (3A, 3B
  \& 3C): System Programming Guide}}.
\newblock   \bibinfo{volume}{253665} (\bibinfo{year}{2016}).
\newblock


\bibitem[\protect\citeauthoryear{Intel}{Intel}{2017}]%
        {Intel_opt}
\bibfield{author}{\bibinfo{person}{Intel}.} \bibinfo{year}{2017}\natexlab{}.
\newblock \bibinfo{title}{{Intel{$\textregistered$} 64 and IA-32 Architectures
  Optimization Reference Manual}}.
\newblock   (\bibinfo{year}{2017}).
\newblock


\bibitem[\protect\citeauthoryear{{Intel Corp.}}{{Intel Corp.}}{2018}]%
        {IntelSpecAnalysis}
\bibfield{author}{\bibinfo{person}{{Intel Corp.}}}
  \bibinfo{year}{2018}\natexlab{}.
\newblock \bibinfo{title}{{Intel} Analysis of Speculative Execution Side
  Channels}.
\newblock
  \bibinfo{howpublished}{\url{https://newsroom.intel.com/wp-content/uploads/sites/11/2018/01/Intel-Analysis-of-Speculative-Execution-Side-Channels.pdf}}.
    (\bibinfo{date}{Jan.} \bibinfo{year}{2018}).
\newblock


\bibitem[\protect\citeauthoryear{{Intel Newsroom}}{{Intel Newsroom}}{2018a}]%
        {Intel2018spectrefix}
\bibfield{author}{\bibinfo{person}{{Intel Newsroom}}.}
  \bibinfo{year}{2018}\natexlab{a}.
\newblock \bibinfo{title}{Advancing Security at the Silicon Level}.
\newblock   (\bibinfo{date}{March} \bibinfo{year}{2018}).
\newblock
\showURL{%
\url{https://newsroom.intel.com/editorials/advancing-security-silicon-level/}}


\bibitem[\protect\citeauthoryear{{Intel Newsroom}}{{Intel Newsroom}}{2018b}]%
        {Intel2018microrevguide}
\bibfield{author}{\bibinfo{person}{{Intel Newsroom}}.}
  \bibinfo{year}{2018}\natexlab{b}.
\newblock \bibinfo{title}{Microcode Revision Guidance}.
\newblock   (\bibinfo{date}{April} \bibinfo{year}{2018}).
\newblock
\showURL{%
\url{https://newsroom.intel.com/wp-content/uploads/sites/11/2018/04/microcode-update-guidance.pdf}}


\bibitem[\protect\citeauthoryear{Irazoqui, Eisenbarth, and Sunar}{Irazoqui
  et~al\mbox{.}}{2015}]%
        {Irazoqui2015SA}
\bibfield{author}{\bibinfo{person}{Gorka Irazoqui}, \bibinfo{person}{Thomas
  Eisenbarth}, {and} \bibinfo{person}{Berk Sunar}.}
  \bibinfo{year}{2015}\natexlab{}.
\newblock \showarticletitle{{S\$A: A Shared Cache Attack that Works Across
  Cores and Defies VM Sandboxing -- and its Application to AES}}. In
  \bibinfo{booktitle}{{\em S\&P'15}}.
\newblock


\bibitem[\protect\citeauthoryear{Irazoqui, Inci, Eisenbarth, and
  Sunar}{Irazoqui et~al\mbox{.}}{2014}]%
        {Irazoqui2014}
\bibfield{author}{\bibinfo{person}{Gorka Irazoqui},
  \bibinfo{person}{Mehmet~Sinan Inci}, \bibinfo{person}{Thomas Eisenbarth},
  {and} \bibinfo{person}{Berk Sunar}.} \bibinfo{year}{2014}\natexlab{}.
\newblock \showarticletitle{{Wait a minute! A fast, Cross-VM attack on AES}}.
  In \bibinfo{booktitle}{{\em RAID'14}}.
\newblock


\bibitem[\protect\citeauthoryear{Jang, Lee, and Kim}{Jang
  et~al\mbox{.}}{2016}]%
        {Jang2016}
\bibfield{author}{\bibinfo{person}{Yeongjin Jang}, \bibinfo{person}{Sangho
  Lee}, {and} \bibinfo{person}{Taesoo Kim}.} \bibinfo{year}{2016}\natexlab{}.
\newblock \showarticletitle{{Breaking Kernel Address Space Layout Randomization
  with Intel TSX}}. In \bibinfo{booktitle}{{\em CCS}}.
\newblock


\bibitem[\protect\citeauthoryear{Jayasinghe, Fernando, Herath, and
  Ragel}{Jayasinghe et~al\mbox{.}}{2010}]%
        {Jayasinghe2010remote}
\bibfield{author}{\bibinfo{person}{Darshana Jayasinghe},
  \bibinfo{person}{Jayani Fernando}, \bibinfo{person}{Ranil Herath}, {and}
  \bibinfo{person}{Roshan Ragel}.} \bibinfo{year}{2010}\natexlab{}.
\newblock \showarticletitle{Remote cache timing attack on advanced encryption
  standard and countermeasures}. In \bibinfo{booktitle}{{\em 5th International
  Conference on Information and Automation for Sustainability (ICIAFs)}}.
\newblock


\bibitem[\protect\citeauthoryear{Kocher}{Kocher}{2018}]%
        {Kocher2018mitigations}
\bibfield{author}{\bibinfo{person}{Paul Kocher}.}
  \bibinfo{year}{2018}\natexlab{}.
\newblock \bibinfo{title}{Spectre Mitigations in Microsoft's C/C++ Compiler}.
\newblock   (\bibinfo{year}{2018}).
\newblock
\showURL{%
\url{https://www.paulkocher.com/doc/MicrosoftCompilerSpectreMitigation.html}}


\bibitem[\protect\citeauthoryear{Kocher, Genkin, Gruss, Haas, Hamburg, Lipp,
  Mangard, Prescher, Schwarz, and Yarom}{Kocher et~al\mbox{.}}{2018}]%
        {Kocher2018}
\bibfield{author}{\bibinfo{person}{Paul Kocher}, \bibinfo{person}{Daniel
  Genkin}, \bibinfo{person}{Daniel Gruss}, \bibinfo{person}{Werner Haas},
  \bibinfo{person}{Mike Hamburg}, \bibinfo{person}{Moritz Lipp},
  \bibinfo{person}{Stefan Mangard}, \bibinfo{person}{Thomas Prescher},
  \bibinfo{person}{Michael Schwarz}, {and} \bibinfo{person}{Yuval Yarom}.}
  \bibinfo{year}{2018}\natexlab{}.
\newblock \showarticletitle{{Spectre Attacks: Exploiting Speculative
  Execution}}.
\newblock \bibinfo{journal}{{\em arXiv:1801.01203\/}} (\bibinfo{year}{2018}).
\newblock


\bibitem[\protect\citeauthoryear{Kocher}{Kocher}{1996}]%
        {Kocher1996}
\bibfield{author}{\bibinfo{person}{Paul~C. Kocher}.}
  \bibinfo{year}{1996}\natexlab{}.
\newblock \showarticletitle{{Timing Attacks on Implementations of
  Diffe-Hellman, RSA, DSS, and Other Systems}}. In \bibinfo{booktitle}{{\em
  CRYPTO}}.
\newblock


\bibitem[\protect\citeauthoryear{Kushner}{Kushner}{2013}]%
        {Kushner2013real}
\bibfield{author}{\bibinfo{person}{David Kushner}.}
  \bibinfo{year}{2013}\natexlab{}.
\newblock \showarticletitle{The real story of stuxnet}.
\newblock \bibinfo{journal}{{\em {IEEE} Spectrum\/}} \bibinfo{volume}{50},
  \bibinfo{number}{3} (\bibinfo{year}{2013}), \bibinfo{pages}{48--53}.
\newblock


\bibitem[\protect\citeauthoryear{Langner}{Langner}{2011}]%
        {Langner2011stuxnet}
\bibfield{author}{\bibinfo{person}{Ralph Langner}.}
  \bibinfo{year}{2011}\natexlab{}.
\newblock \showarticletitle{Stuxnet: Dissecting a cyberwarfare weapon}.
\newblock \bibinfo{journal}{{\em IEEE Security \& Privacy\/}}
  \bibinfo{volume}{9}, \bibinfo{number}{3} (\bibinfo{year}{2011}),
  \bibinfo{pages}{49--51}.
\newblock


\bibitem[\protect\citeauthoryear{Lawall}{Lawall}{2018}]%
        {LWN2018spectrecoccinelle}
\bibfield{author}{\bibinfo{person}{Julia Lawall}.}
  \bibinfo{year}{2018}\natexlab{}.
\newblock \bibinfo{title}{Re: [RFC PATCH] asm/generic: introduce if\_nospec and
  nospec\_barrier}.
\newblock   (\bibinfo{year}{2018}).
\newblock
\showURL{%
\url{https://lwn.net/Articles/743287/}}


\bibitem[\protect\citeauthoryear{Lee, Shih, Gera, Kim, Kim, and Peinado}{Lee
  et~al\mbox{.}}{2017}]%
        {Lee2017BranchShadowing}
\bibfield{author}{\bibinfo{person}{Sangho Lee}, \bibinfo{person}{Ming{-}Wei
  Shih}, \bibinfo{person}{Prasun Gera}, \bibinfo{person}{Taesoo Kim},
  \bibinfo{person}{Hyesoon Kim}, {and} \bibinfo{person}{Marcus Peinado}.}
  \bibinfo{year}{2017}\natexlab{}.
\newblock \showarticletitle{Inferring Fine-grained Control Flow Inside {SGX}
  Enclaves with Branch Shadowing}. In \bibinfo{booktitle}{{\em {USENIX}
  Security Symposium}}.
\newblock


\bibitem[\protect\citeauthoryear{{Lipp}, {Gruss}, {Spreitzer}, Maurice, and
  {Mangard}}{{Lipp} et~al\mbox{.}}{2016}]%
        {Lipp2016}
\bibfield{author}{\bibinfo{person}{Moritz {Lipp}}, \bibinfo{person}{Daniel
  {Gruss}}, \bibinfo{person}{Raphael {Spreitzer}},
  \bibinfo{person}{Cl\'{e}mentine Maurice}, {and} \bibinfo{person}{Stefan
  {Mangard}}.} \bibinfo{year}{2016}\natexlab{}.
\newblock \showarticletitle{{ARMageddon: Cache Attacks on Mobile Devices}}. In
  \bibinfo{booktitle}{{\em USENIX Security Symposium}}.
\newblock


\bibitem[\protect\citeauthoryear{Lipp, Schwarz, Gruss, Prescher, Haas, Fogh,
  Horn, Mangard, Kocher, Genkin, Yarom, and Hamburg}{Lipp
  et~al\mbox{.}}{2018}]%
        {Lipp2018meltdown}
\bibfield{author}{\bibinfo{person}{Moritz Lipp}, \bibinfo{person}{Michael
  Schwarz}, \bibinfo{person}{Daniel Gruss}, \bibinfo{person}{Thomas Prescher},
  \bibinfo{person}{Werner Haas}, \bibinfo{person}{Anders Fogh},
  \bibinfo{person}{Jann Horn}, \bibinfo{person}{Stefan Mangard},
  \bibinfo{person}{Paul Kocher}, \bibinfo{person}{Daniel Genkin},
  \bibinfo{person}{Yuval Yarom}, {and} \bibinfo{person}{Mike Hamburg}.}
  \bibinfo{year}{2018}\natexlab{}.
\newblock \showarticletitle{Meltdown: Reading Kernel Memory from User Space}.
  In \bibinfo{booktitle}{{\em {USENIX} Security Symposium}}.
\newblock


\bibitem[\protect\citeauthoryear{Liu, Yarom, Ge, Heiser, and Lee}{Liu
  et~al\mbox{.}}{2015}]%
        {Liu2015}
\bibfield{author}{\bibinfo{person}{Fangfei Liu}, \bibinfo{person}{Yuval Yarom},
  \bibinfo{person}{Qian Ge}, \bibinfo{person}{Gernot Heiser}, {and}
  \bibinfo{person}{Ruby~B. Lee}.} \bibinfo{year}{2015}\natexlab{}.
\newblock \showarticletitle{{Last-Level Cache Side-Channel Attacks are
  Practical}}. In \bibinfo{booktitle}{{\em {IEEE} Symposium on Security and
  Privacy -- {SP}}}. \bibinfo{publisher}{{IEEE} Computer Society},
  \bibinfo{pages}{605--622}.
\newblock


\bibitem[\protect\citeauthoryear{Liu, Gao, and Reiter}{Liu
  et~al\mbox{.}}{2017}]%
        {Liu2017demand}
\bibfield{author}{\bibinfo{person}{Weijie Liu}, \bibinfo{person}{Debin Gao},
  {and} \bibinfo{person}{Michael~K Reiter}.} \bibinfo{year}{2017}\natexlab{}.
\newblock \showarticletitle{On-demand time blurring to support side-channel
  defense}. In \bibinfo{booktitle}{{\em ESORICS}}.
\newblock


\bibitem[\protect\citeauthoryear{Maisuradze and Rossow}{Maisuradze and
  Rossow}{2018}]%
        {Maisuradze2018speculose}
\bibfield{author}{\bibinfo{person}{Giorgi Maisuradze} {and}
  \bibinfo{person}{Christian Rossow}.} \bibinfo{year}{2018}\natexlab{}.
\newblock \showarticletitle{Speculose: Analyzing the Security Implications of
  Speculative Execution in CPUs}.
\newblock \bibinfo{journal}{{\em arXiv:1801.04084\/}} (\bibinfo{year}{2018}).
\newblock


\bibitem[\protect\citeauthoryear{Marco-Gisbert and Ripoll-Ripoll}{Marco-Gisbert
  and Ripoll-Ripoll}{2016}]%
        {Marco2016exploiting}
\bibfield{author}{\bibinfo{person}{Hector Marco-Gisbert} {and}
  \bibinfo{person}{Ismael Ripoll-Ripoll}.} \bibinfo{year}{2016}\natexlab{}.
\newblock \showarticletitle{Exploiting Linux and PaX ASLR’s weaknesses on
  32-and 64-bit systems}.
\newblock \bibinfo{journal}{{\em BlackHat Asia\/}} (\bibinfo{year}{2016}).
\newblock


\bibitem[\protect\citeauthoryear{Maurice, Neumann, Heen, and
  Francillon}{Maurice et~al\mbox{.}}{2015}]%
        {Maurice2015C5}
\bibfield{author}{\bibinfo{person}{Cl\'{e}mentine Maurice},
  \bibinfo{person}{Christoph Neumann}, \bibinfo{person}{Olivier Heen}, {and}
  \bibinfo{person}{Aur\'{e}lien Francillon}.} \bibinfo{year}{2015}\natexlab{}.
\newblock \showarticletitle{{C5: Cross-Cores Cache Covert Channel}}. In
  \bibinfo{booktitle}{{\em DIMVA}}.
\newblock


\bibitem[\protect\citeauthoryear{Maurice, Weber, Schwarz, Giner, Gruss,
  Alberto~Boano, Mangard, and Römer}{Maurice et~al\mbox{.}}{2017}]%
        {Maurice2017Hello}
\bibfield{author}{\bibinfo{person}{Clémentine Maurice},
  \bibinfo{person}{Manuel Weber}, \bibinfo{person}{Michael Schwarz},
  \bibinfo{person}{Lukas Giner}, \bibinfo{person}{Daniel Gruss},
  \bibinfo{person}{Carlo Alberto~Boano}, \bibinfo{person}{Stefan Mangard},
  {and} \bibinfo{person}{Kay Römer}.} \bibinfo{year}{2017}\natexlab{}.
\newblock \showarticletitle{{Hello from the Other Side: SSH over Robust Cache
  Covert Channels in the Cloud}}. In \bibinfo{booktitle}{{\em NDSS}}.
\newblock


\bibitem[\protect\citeauthoryear{McCalpin}{McCalpin}{2015}]%
        {Fog2015AVXa}
\bibfield{author}{\bibinfo{person}{John~D. McCalpin}.}
  \bibinfo{year}{2015}\natexlab{}.
\newblock \bibinfo{title}{Test results for Intel's Sandy Bridge processor}.
\newblock   (\bibinfo{year}{2015}).
\newblock
\showURL{%
\url{http://agner.org/optimize/blog/read.php?i=378#378}}


\bibitem[\protect\citeauthoryear{Oberman, Favor, and Weber}{Oberman
  et~al\mbox{.}}{1999}]%
        {Oberman1999amd}
\bibfield{author}{\bibinfo{person}{Stuart Oberman}, \bibinfo{person}{Greg
  Favor}, {and} \bibinfo{person}{Fred Weber}.} \bibinfo{year}{1999}\natexlab{}.
\newblock \showarticletitle{AMD 3DNow! technology: Architecture and
  implementations}.
\newblock \bibinfo{journal}{{\em IEEE Micro\/}} \bibinfo{volume}{19},
  \bibinfo{number}{2} (\bibinfo{year}{1999}), \bibinfo{pages}{37--48}.
\newblock


\bibitem[\protect\citeauthoryear{Osvik, Shamir, and Tromer}{Osvik
  et~al\mbox{.}}{2006}]%
        {Osvik2006}
\bibfield{author}{\bibinfo{person}{Dag~Arne Osvik}, \bibinfo{person}{Adi
  Shamir}, {and} \bibinfo{person}{Eran Tromer}.}
  \bibinfo{year}{2006}\natexlab{}.
\newblock \showarticletitle{{Cache Attacks and Countermeasures: the Case of
  AES}}. In \bibinfo{booktitle}{{\em CT-RSA}}.
\newblock


\bibitem[\protect\citeauthoryear{Pardoe}{Pardoe}{2018}]%
        {Microsoft2018Spectre}
\bibfield{author}{\bibinfo{person}{Andrew Pardoe}.}
  \bibinfo{year}{2018}\natexlab{}.
\newblock \bibinfo{title}{Spectre mitigations in MSVC}.
\newblock   (\bibinfo{year}{2018}).
\newblock
\showURL{%
\url{https://blogs.msdn.microsoft.com/vcblog/2018/01/15/spectre-mitigations-in-msvc/}}


\bibitem[\protect\citeauthoryear{{PaX Team}}{{PaX Team}}{2003}]%
        {PaxASLR}
\bibfield{author}{\bibinfo{person}{{PaX Team}}.}
  \bibinfo{year}{2003}\natexlab{}.
\newblock \bibinfo{title}{Address space layout randomization (ASLR)}.
\newblock   (\bibinfo{year}{2003}).
\newblock
\showURL{%
\url{http://pax.grsecurity.net/docs/aslr.txt}}


\bibitem[\protect\citeauthoryear{Peleg and Weiser}{Peleg and Weiser}{1996}]%
        {Peleg1996mmx}
\bibfield{author}{\bibinfo{person}{Alex Peleg} {and} \bibinfo{person}{Uri
  Weiser}.} \bibinfo{year}{1996}\natexlab{}.
\newblock \showarticletitle{MMX technology extension to the Intel
  architecture}.
\newblock \bibinfo{journal}{{\em {IEEE} Micro\/}} \bibinfo{volume}{16},
  \bibinfo{number}{4} (\bibinfo{year}{1996}), \bibinfo{pages}{42--50}.
\newblock


\bibitem[\protect\citeauthoryear{Percival}{Percival}{2005}]%
        {Percival2005}
\bibfield{author}{\bibinfo{person}{Colin Percival}.}
  \bibinfo{year}{2005}\natexlab{}.
\newblock \showarticletitle{{Cache missing for fun and profit}}. In
  \bibinfo{booktitle}{{\em Proceedings of BSDCan}}.
\newblock


\bibitem[\protect\citeauthoryear{Pessl, Gruss, Maurice, Schwarz, and
  Mangard}{Pessl et~al\mbox{.}}{2016}]%
        {Pessl2016}
\bibfield{author}{\bibinfo{person}{Peter Pessl}, \bibinfo{person}{Daniel
  Gruss}, \bibinfo{person}{Cl\'{e}mentine Maurice}, \bibinfo{person}{Michael
  Schwarz}, {and} \bibinfo{person}{Stefan Mangard}.}
  \bibinfo{year}{2016}\natexlab{}.
\newblock \showarticletitle{{DRAMA: Exploiting DRAM Addressing for Cross-CPU
  Attacks}}. In \bibinfo{booktitle}{{\em USENIX Security Symposium}}.
\newblock


\bibitem[\protect\citeauthoryear{Pukelsheim}{Pukelsheim}{1994}]%
        {Pukelsheim1994}
\bibfield{author}{\bibinfo{person}{Friedrich Pukelsheim}.}
  \bibinfo{year}{1994}\natexlab{}.
\newblock \showarticletitle{The three sigma rule}.
\newblock \bibinfo{journal}{{\em The American Statistician\/}}
  (\bibinfo{year}{1994}).
\newblock


\bibitem[\protect\citeauthoryear{Schwarz, Gruss, Lipp, Maurice, Schuster, Fogh,
  and Mangard}{Schwarz et~al\mbox{.}}{2018}]%
        {Schwarz2018DF}
\bibfield{author}{\bibinfo{person}{Michael Schwarz}, \bibinfo{person}{Daniel
  Gruss}, \bibinfo{person}{Moritz Lipp}, \bibinfo{person}{Cl{\'e}mentine
  Maurice}, \bibinfo{person}{Thomas Schuster}, \bibinfo{person}{Anders Fogh},
  {and} \bibinfo{person}{Stefan Mangard}.} \bibinfo{year}{2018}\natexlab{}.
\newblock \showarticletitle{Automated Detection, Exploitation, and Elimination
  of Double-Fetch Bugs using Modern CPU Features}.
\newblock \bibinfo{journal}{{\em AsiaCCS\/}} (\bibinfo{year}{2018}).
\newblock


\bibitem[\protect\citeauthoryear{Schwarz, Gruss, Weiser, Maurice, and
  Mangard}{Schwarz et~al\mbox{.}}{2017}]%
        {Schwarz2017MGX}
\bibfield{author}{\bibinfo{person}{Michael Schwarz}, \bibinfo{person}{Daniel
  Gruss}, \bibinfo{person}{Samuel Weiser}, \bibinfo{person}{Clémentine
  Maurice}, {and} \bibinfo{person}{Stefan Mangard}.}
  \bibinfo{year}{2017}\natexlab{}.
\newblock \showarticletitle{{Malware Guard Extension: Using SGX to Conceal
  Cache Attacks }}. In \bibinfo{booktitle}{{\em DIMVA'17}}.
\newblock


\bibitem[\protect\citeauthoryear{Schwarz, Lipp, Gruss, Weiser, Maurice,
  Spreitzer, and Mangard}{Schwarz et~al\mbox{.}}{2018}]%
        {Schwarz2018KeyDrown}
\bibfield{author}{\bibinfo{person}{Michael Schwarz}, \bibinfo{person}{Moritz
  Lipp}, \bibinfo{person}{Daniel Gruss}, \bibinfo{person}{Samuel Weiser},
  \bibinfo{person}{Clémentine Maurice}, \bibinfo{person}{Raphael Spreitzer},
  {and} \bibinfo{person}{Stefan Mangard}.} \bibinfo{year}{2018}\natexlab{}.
\newblock \showarticletitle{{KeyDrown: Eliminating Software-Based Keystroke
  Timing Side-Channel Attacks}}. In \bibinfo{booktitle}{{\em NDSS}}.
\newblock


\bibitem[\protect\citeauthoryear{Tankard}{Tankard}{2011}]%
        {Tankard2011advanced}
\bibfield{author}{\bibinfo{person}{Colin Tankard}.}
  \bibinfo{year}{2011}\natexlab{}.
\newblock \showarticletitle{Advanced persistent threats and how to monitor and
  deter them}.
\newblock \bibinfo{journal}{{\em Network security\/}} \bibinfo{volume}{2011},
  \bibinfo{number}{8} (\bibinfo{year}{2011}), \bibinfo{pages}{16--19}.
\newblock


\bibitem[\protect\citeauthoryear{Trippel, Lustig, and Martonosi}{Trippel
  et~al\mbox{.}}{2018}]%
        {Trippel2018MeltdownPrime}
\bibfield{author}{\bibinfo{person}{Caroline Trippel}, \bibinfo{person}{Daniel
  Lustig}, {and} \bibinfo{person}{Margaret Martonosi}.}
  \bibinfo{year}{2018}\natexlab{}.
\newblock \showarticletitle{MeltdownPrime and SpectrePrime:
  Automatically-Synthesized Attacks Exploiting Invalidation-Based Coherence
  Protocols}.
\newblock \bibinfo{journal}{{\em arXiv:1802.03802\/}} (\bibinfo{year}{2018}).
\newblock


\bibitem[\protect\citeauthoryear{Tsunoo, Saito, and Suzaki}{Tsunoo
  et~al\mbox{.}}{2003}]%
        {Tsunoo2003}
\bibfield{author}{\bibinfo{person}{Yukiyasu Tsunoo}, \bibinfo{person}{Teruo
  Saito}, {and} \bibinfo{person}{Tomoyasu Suzaki}.}
  \bibinfo{year}{2003}\natexlab{}.
\newblock \showarticletitle{{Cryptanalysis of DES implemented on computers with
  cache}}. In \bibinfo{booktitle}{{\em {CHES'03}}}. \bibinfo{pages}{62--76}.
\newblock


\bibitem[\protect\citeauthoryear{Turner}{Turner}{2018}]%
        {Turner2018retpoline}
\bibfield{author}{\bibinfo{person}{Paul Turner}.}
  \bibinfo{year}{2018}\natexlab{}.
\newblock \bibinfo{title}{Retpoline: a software construct for preventing
  branch-target-injection}.
\newblock   (\bibinfo{year}{2018}).
\newblock


\bibitem[\protect\citeauthoryear{Weimer}{Weimer}{2018}]%
        {GCC2018retpolinecfi}
\bibfield{author}{\bibinfo{person}{Florian Weimer}.}
  \bibinfo{year}{2018}\natexlab{}.
\newblock \bibinfo{title}{Retpolines and CFI}.
\newblock   (\bibinfo{year}{2018}).
\newblock
\showURL{%
\url{https://gcc.gnu.org/ml/gcc/2018-01/msg00160.html}}


\bibitem[\protect\citeauthoryear{Yarom and Falkner}{Yarom and Falkner}{2014}]%
        {Yarom2014}
\bibfield{author}{\bibinfo{person}{Yuval Yarom} {and} \bibinfo{person}{Katrina
  Falkner}.} \bibinfo{year}{2014}\natexlab{}.
\newblock \showarticletitle{{Flush+Reload: a High Resolution, Low Noise, L3
  Cache Side-Channel Attack}}. In \bibinfo{booktitle}{{\em USENIX Security
  Symposium}}.
\newblock


\bibitem[\protect\citeauthoryear{Zhang, Xiao, and Zhang}{Zhang
  et~al\mbox{.}}{2016}]%
        {Zhang2016ROPFR}
\bibfield{author}{\bibinfo{person}{Xiaokuan Zhang}, \bibinfo{person}{Yuan
  Xiao}, {and} \bibinfo{person}{Yinqian Zhang}.}
  \bibinfo{year}{2016}\natexlab{}.
\newblock \showarticletitle{Return-oriented flush-reload side channels on arm
  and their implications for android devices}. In \bibinfo{booktitle}{{\em
  Proceedings of the 2016 ACM SIGSAC Conference on Computer and Communications
  Security}}. \bibinfo{pages}{858--870}.
\newblock


\bibitem[\protect\citeauthoryear{Zhang, Juels, Reiter, and Ristenpart}{Zhang
  et~al\mbox{.}}{2014}]%
        {Zhang2014}
\bibfield{author}{\bibinfo{person}{Yinqian Zhang}, \bibinfo{person}{Ari Juels},
  \bibinfo{person}{Michael~K. Reiter}, {and} \bibinfo{person}{Thomas
  Ristenpart}.} \bibinfo{year}{2014}\natexlab{}.
\newblock \showarticletitle{{Cross-Tenant Side-Channel Attacks in PaaS
  Clouds}}. In \bibinfo{booktitle}{{\em CCS'14}}.
\newblock


\bibitem[\protect\citeauthoryear{Zhao, Wang, and Zheng}{Zhao
  et~al\mbox{.}}{2009}]%
        {Zhao2009cache}
\bibfield{author}{\bibinfo{person}{Xin-jie Zhao}, \bibinfo{person}{Tao Wang},
  {and} \bibinfo{person}{Yuanyuan Zheng}.} \bibinfo{year}{2009}\natexlab{}.
\newblock \showarticletitle{Cache Timing Attacks on Camellia Block Cipher.}
\newblock \bibinfo{howpublished}{Cryptology ePrint Archive, Report 2009/354}.
\newblock  (\bibinfo{year}{2009}).
\newblock


\end{thebibliography}
%
%

\end{document}